\documentclass[preprint,12pt]{elsarticle}

 \usepackage{graphics}
 \usepackage{epsfig}

\usepackage{amssymb}

 \usepackage{graphics}
 \usepackage{epsfig}

\usepackage{amssymb}

\newcommand{\bra}[1]{\langle #1|}	
\newcommand{\ket}[1]{|#1\rangle}
\newcommand{\braket}[2]{\langle #1|#2\rangle}
\usepackage{amsmath}
\usepackage{amssymb}
\usepackage{dsfont}
\usepackage{pst-all}

\journal{Chemical Physics}

\begin{document}

\begin{frontmatter}


\title{The dissipative quantum Duffing oscillator: a comparison of Floquet-based approaches}


\author{Carmen Vierheilig and Milena Grifoni}

\address{Institut f\"ur Theoretische Physik, Universit\"at Regensburg}

\begin{abstract}
We study the dissipative quantum Duffing oscillator in the deep quantum regime with two different approaches: The first is based on the exact Floquet states of the linear oscillator and the nonlinearity is treated perturbatively. It well describes the nonlinear oscillator dynamics away from resonance. The second, in contrast, is applicable at and in the vicinity of a $N$-photon resonance and it exploits quasi-degenerate perturbation theory for the nonlinear oscillator in Floquet space. It is perturbative both in driving and nonlinearity. A combination of both approaches yields the possibility to cover the whole range of driving frequencies. As an example we discuss the dissipative dynamics of the Duffing oscillator near and at the one-photon resonance.

\end{abstract}


\end{frontmatter}

\section{Introduction}
Classical nonlinear systems show interesting phenomena like bistability, frequency doubling and nonlinear response and cover a wide range of applicability and various physical realizations \cite{Nayfeh,Jordan}. In the last years it has been possible to build nonlinear devices which can potentially reach the quantum regime. These are for example cavities incorporating a Josephson junction \cite{Boaknin,Metcalfe}, SQUIDs used as bifurcation amplifiers to improve qubit read-out \cite{Lee,Picot,Siddiqi,Siddiqi2} or nanomechanical resonators \cite{Almog,Alridge}. 
Recently, a novel class of devices combining SQUIDS and resonators has been demonstrated. For example, sensitive detection of the position of a  micromechanical resonator embedded in a nonlinear, strongly damped DC-SQUID has been achieved \cite{Etaki}. In the deep quantum regime, where bistability is no longer observed, there has been to date no experimental operation of nonlinear driven oscillators to our knowledge.\\
From the theoretical side, semiclassical approaches have been used to describe the situation where the underlying classical bistability still plays a dominant role. A DC-SQUID embedded in a cavity allowing displacement detection and cooling was analyzed in \cite{Nation}. 
Composed qubit-Josephson bifurcation amplifier systems have been considered in \cite{SerbanDet,Serbanqubit}. Dynamical tunneling in a Duffing oscillator was accounted for in \cite{SerbanMQDT} within a semiclassical WKB scheme, while in \cite{Guo,Katz1,Katz2} a Wigner function analysis near the bifurcation point is put forward.\\
The behaviour of the Duffing oscillator (DO) in the deep quantum regime has attracted recently lot of interest. In particular Rigo et al. \cite{Rigo} demonstrated, based on a quantum diffusion model, that in the steady state the quantum DO does not exhibit any bistability or hysteresis. It was also shown that  
the response of the Duffing oscillator displays antiresonant dips and resonant peaks \cite{Fistul,Peano1,Peano2,Peano3} depending on the frequency of the driving field, originating from special degeneracies of the eigenenergy spectrum of the nonlinear oscillator \cite{Fistul}.
While the antiresonances persist in the presence of a weak Ohmic bath, for strong damping the nonlinear response turns to a resonant behaviour, namely the one of a linear oscillator at a shifted frequency \cite{Peano2}. Finally, recently Nakano et al. \cite{Nakano} looked at the composed qubit-DO dynamics during read-out process.\\
In this work we investigate the deep quantum limit of the quantum Duffing oscillator and present two different approaches covering different parameter regimes. The first approach is based on the exact Floquet energies and states of the driven linear oscillator with the nonlinearity treated perturbatively. As there is no restriction on the driving amplitude, this scheme can also be applied to the regime where the driving amplitude is larger than the nonlinearity. The second approach treats both the driving and the nonlinearity perturbatively. It is applicable for driving frequencies which can resonantly excite two states of the nonlinear oscillator, requiring that the driving cannot overcome the nonlinearity. In general a combination of both approaches allows to cover the whole range of driving frequencies. Exemplarily we consider the dynamics of the Duffing oscillator near the one-photon resonance, where the oscillator dynamics is described analytically. As in \cite{Fistul,Peano1,Peano2,Peano3} we obtain that for weak dissipation the amplitude of the oscillations displays an antiresonance rather than a resonance. We find a characteristic asymmetry of the antiresonance lineshape. In contrast to \cite{Fistul,Peano1,Peano2,Peano3}, our analytic results are obtained without applying a rotating wave approximation (RWA) on the Duffing oscillator.\\
The paper is organized as follows: In section \ref{QDOmodel} we introduce the Hamiltonian of the non-dissipative Duffing oscillator and the two Floquet based approximation schemes to treat it. The energy spectrum and eigenstates of the non-dissipative system are calculated with the two different schemes in section \ref{GFS} and section \ref{PertApp} and both approaches are compared in section \ref{Comparison}. Afterwards dissipative effects are included within a Born-Markov-Floquet master equation in section \ref{Dissipation}. Section \ref{Observable} addresses the special case of the one-photon resonance including dissipative effects. In section \ref{Conclusions} conclusions are drawn.

\section{Quantum Duffing oscillator}\label{QDOmodel}
A quantum Duffing oscillator is described by the Hamiltonian:
\begin{eqnarray}
\hat{H}_{{\rm DO}}(t)=\frac{\hat{P}_y^2}{2M}+\frac{M\Omega^2}{2}\hat{y}^2+\frac{\alpha}{4}\hat{y}^4+\hat{y}F\cos(\omega_{ex} t), 
\end{eqnarray}
where $M$ and $\Omega$ are the mass and frequency of the Duffing oscillator which is driven by a monochromatic field of amplitude $F$ and frequency $\omega_{ex}$. For later convenience we introduce the oscillator length $y_0:=\sqrt{\frac{\hbar}{M\Omega}}$.
In the following we will consider the case of hard nonlinearities, $\alpha>0$, such that the undriven potential is monostable.\\
To treat the quantum Duffing oscillator problem we observe that the Hamiltonian can be rewritten as:
\begin{subequations}
\begin{eqnarray}
\hat{H}_{{\rm DO}}(t)&=&\hat{H}_{{\rm LO}}(t)+\frac{\alpha}{4}\hat{y}^4\label{App1}\\
&=&\hat{H}_{{\rm NLO}}+\hat{y}F\cos(\omega_{ex}t),\label{App2}
\end{eqnarray}
\end{subequations}
where $\hat{H}_{{\rm LO}}(t)$ describes a driven linear oscillator, while $\hat{H}_{{\rm NLO}}$ is the Hamiltonian of an undriven nonlinear oscillator. Due to the periodic driving Floquet theory \cite{Floquet,Shirley,Sambe,Grifoni304}, reviewed in \ref{FloqTh}, can be applied. In particular Floquet theorem states that solutions of the time-dependent Schr\"odinger equation for $\hat{H}_{{\rm DO}}(t)$ are of the form
\begin{equation}
\ket{\psi_j(t)}=\exp(-i\epsilon_j t/\hbar)|\phi_j(t)\rangle,
\end{equation}
where $|\phi_j(t)\rangle=|\phi_j(t+T_{\omega_{ex}})\rangle$. The quasienergies $\epsilon_j$ and Floquet states $|\phi_j(t)\rangle$, respectively, are eigenvalues and eigenfunctions of the Floquet Hamiltonian $\hat{\mathcal{H}}_{\rm DO}(t)=\hat{H}_{\rm DO}(t)-i\hbar\partial_t$.
As discussed in \ref{FloqTh}, $|\phi_{j,n}(t)\rangle=|\phi_j(t)\rangle\exp(-in\omega_{ex}t)$ yields a physically equivalent solution but with shifted quasienergy $\epsilon_{j,n}=\epsilon_j-n\hbar\omega_{ex}$.
Eqs. (\ref{App1}) and (\ref{App2}) suggest two different approaches, shown in Figure \ref{Schaubild}, to solve the eigenvalue problem described by Eqs. (\ref{heigenvalue}) and (\ref{fouriereigenvalue}):
\begin{figure}[h!]
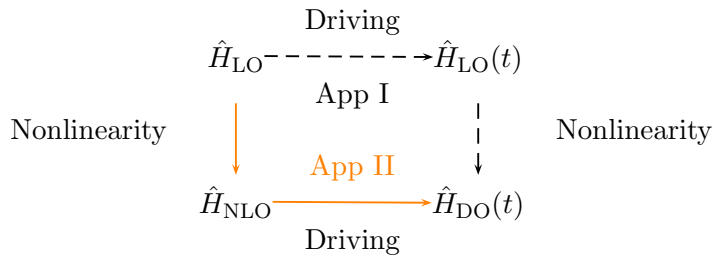

  \centering
  \psset{linewidth=.25mm} 
  \small{
  \begin{tabular}{ccccc}
& &Driving & &\\
 &\rnode{7}{$\hat{H}_{\rm LO}$}  &\rnode{3}{ }   &\rnode{4}{$\hat{H}_{\rm LO}(t)$}& \\
&\rnode{2}{}  &{\rm App I} & \rnode{8}{}&\\
Nonlinearity& & & &Nonlinearity\\
&\rnode{9}{} &{\rm  \textcolor{orange}{App II}}& \rnode{10}{}&\\
 & \rnode{5}{$\hat{H}_{\rm NLO}$}& & \rnode{6}{$\hat{H}_{\rm DO}(t)$}&\\
 & &Driving & & \\

  \end{tabular}}
  \ncline[linestyle=dashed]{->}{7}{4}
  \ncline[linecolor=orange]{->}{2}{9}
  \ncline[linestyle=dashed]{->}{8}{10}
  \ncline[linecolor=orange]{->}{5}{6}
\caption{Different procedures to incorporate driving and nonlinearity. In App I starting point are the exact Floquet states and eigenenergies of the driven linear oscillator $\hat{H}_{LO}(t)$. The nonlinearity is the perturbation. In App II the driving is a perturbation expressed on the basis of the Floquet states of the undriven nonlinear oscillator $\hat{H}_{NLO}$.\label{Schaubild}}
\end{figure}
In the first one, called App I, starting point are the exact Floquet states and eigenenergies of the driven linear oscillator $\hat{H}_{\rm LO}(t)$, see Eq. (\ref{App1}). The nonlinearity is treated as a perturbation. A similar problem was considered by Tittonen et al. \cite{Tittonen}. This approach is convenient if the Floquet states of the time-dependent Hamiltonian are known.\\
For the driven harmonic oscillator they have been derived by Husimi and Perelomov \cite{Husimi,Popov} and are given in \ref{FloqLO}.\\
In the second approach, which we call App II, one considers as unperturbed system the undriven nonlinear oscillator (NLO) and the driving is the perturbation, see Eq. (\ref{App2}).\\
As we shall see, the different ways of treating the infinite dimensional Floquet Hamiltonian result in crucial differences when evaluating observables of the Duffing oscillator.\\
\section{Perturbation theory for a time-periodic Hamiltonian with time-independent perturbation}
\label{GFS}
The starting point of the perturbative treatment App I is the Floquet equation for the full Floquet Hamiltonian $\hat{\mathcal{H}}$ in the extended Hilbert space $\mathcal{R}\otimes\mathcal{T}$, see Eq. (\ref{heigenvalue}),
\begin{eqnarray}\label{FloqEq}
\hat{\mathcal{H}}\ket{\phi_{j,m}}\rangle&=&\epsilon_{j,m}\ket{\phi_{j,m}}\rangle,
\end{eqnarray}
where $\hat{\mathcal{H}}=\hat{\mathcal{H}}_0+\hat{V}_\alpha$. Moreover, the Floquet states of the Floquet Hamiltonian $\hat{\mathcal{H}}_0$ satisfying the eigenvalue equation (\ref{heigenvalue}) are known, see e.g. Eq. (\ref{flo1a}) and Eq. (\ref{flo5}):
\begin{eqnarray}
\hat{\mathcal{H}}_0\ket{\phi_{j,m}}\rangle_0&=&\epsilon_{j,m}^{(0)}\ket{\phi_{j,m}}\rangle_0.
\end{eqnarray} We look for an expression of $\epsilon_{j,m}$ and $\ket{\phi_{j,m}}\rangle$ in first order in $\hat{V}_\alpha$. 
Hence we introduce the first order corrections $\epsilon_{j,m}^{(1)}$ and $|\phi_{j,m}\rangle\rangle_1$ as:
\begin{eqnarray}\label{pert}
\epsilon_{j,m}&=&\epsilon_{j,m}^{(0)}+\epsilon_{j,m}^{(1)},\\
|\phi_{j,m}\rangle\rangle&=&|\phi_{j,m}\rangle\rangle_0+|\phi_{j,m}\rangle\rangle_1\nonumber.
\end{eqnarray}
Because the perturbation is time-independent, it is diagonal in the Hilbert space $\mathcal{T}$. Additionallly, we introduce the Fourier coefficients:
\begin{eqnarray}\label{mefourier}
_0\langle\bra{\phi_{k,n}}\hat{V}_\alpha\ket{\phi_{j,m}}\rangle_0&\equiv&v_{kj}^{(n-m)}.
\end{eqnarray}
As in the case of conventional stationary perturbation theory, the perturbed states are written as a linear combination of the unperturbed states:
\begin{eqnarray}\label{GFSstates}
|\phi_{j,m}\rangle\rangle
&=&|\phi_{j,m}\rangle\rangle_0+\sum_{(i,n)\neq(j,m)} c_{ij}^{nm}|\phi_{i,n}\rangle\rangle_0,
\end{eqnarray}
where $(i,n)$ denotes the couple of quantum numbers $i$ and $n$.
Inserting ansatz Eq. (\ref{pert}) in the Floquet equation (\ref{FloqEq}) we obtain:
\begin{eqnarray}
\left(\hat{\mathcal{H}}_0+\hat{V}_\alpha-\epsilon_{j,m}^{(0)}-\epsilon_{j,m}^{(1)}\right)\left(|\phi_{j,m}\rangle\rangle_0+|\phi_{j,m}\rangle\rangle_1\right)&=&0.
\end{eqnarray}
Because $\epsilon_{j,m}^{(0)}$ and $|\phi_{j,m}\rangle\rangle_0$ solve the Floquet equation for $\hat{\mathcal{H}}_{0}$
the last equation reduces to:
\begin{eqnarray}\label{gfs2}
\left(\hat{V}_\alpha-\epsilon_{j,m}^{(1)}\right)|\phi_{j,m}\rangle\rangle_0+\left(\hat{\mathcal{H}}_{0}-\epsilon_{j,m}^{(0)}\right)|\phi_{j,m}\rangle\rangle_1&=&0.
\end{eqnarray}
This equation allows to determine the modification to the quasienergy $\epsilon_{j,m}^{(1)}$ and the actual form of the coefficients $c_{ij}^{nm}$. To calculate $\epsilon_{j,m}^{(1)}$ we multiply it from the left with $_0\langle\langle\phi_{j,m}|$.
It follows:
\begin{eqnarray}
\epsilon_{j,m}^{(1)}&=&_0\langle\langle\phi_{j,m}|\hat{V}_\alpha|\phi_{j,m}\rangle\rangle_0.
\end{eqnarray}
To determine the coefficients for the states we multiply Eq. (\ref{gfs2}) from the left with $_0\langle\langle\phi_{k,n}|$, where the couple $(k,n)\neq (j,m)$. Moreover we exclude the case of degenerate quasienergies and impose: $\epsilon_{k,n}^{(0)}\neq\epsilon_{j,m}^{(0)}$. In case of $\epsilon_{k,n}^{(0)}=\epsilon_{j,m}^{(0)}$ degenerate perturbation theory should be applied. We obtain:
\begin{eqnarray}
0&=&_0\langle\langle\phi_{k,n}|\hat{V}_\alpha|\phi_{j,m}\rangle_0+(\epsilon_{k,n}^{(0)}-\epsilon_ {j,m}^{(0)})c_{kj}^{nm},
\end{eqnarray}
yielding:
\begin{eqnarray}\label{ccoeff}
c_{kj}^{nm}&=&\frac{_0\langle\langle\phi_{k,n}|\hat{V}_\alpha|\phi_{j,m}\rangle\rangle_0}{\epsilon_{j,m}^{(0)}-\epsilon_{k,n}^{(0)}}\equiv c_{kj}^{(n-m)}.
\end{eqnarray}
If we set the driving to zero the quasienergy and the states reduce to the ones obtained by applying conventional stationary perturbation theory on the unforced system.
\subsection{Application to the quantum Duffing oscillator}
We can now determine the actual form of the quasienergy spectrum and the corresponding expansion coefficients for the case of the quantum Duffing oscillator using as perturbation $\hat{V}_\alpha=\frac{\alpha}{4}\hat{y}^4$. The matrix elements Eq. (\ref{mefourier}) defined as:
\begin{eqnarray}
v_{kj}^{(n-m)}&=&\frac{1}{T_{\omega_{ex}}}\int_0^{T_{\omega_{ex}}}dt\exp(i(n-m)\omega_{ex}t)_0\langle\phi_k(t)|\hat{V}_\alpha|\phi_j(t)\rangle_0,
            \end{eqnarray}
are given in \ref{Fouriercomponents}.
The quasienergies of the quantum Duffing oscillator, exact in all orders of the driving strength and up to first order in the nonlinearity, read:
\begin{eqnarray}\label{GFSenergy}
\epsilon_{j,m}&=&\hbar\Omega\left(j+\frac{1}{2}\right)+\frac{F^2}{4M(\omega_{ex}^2-\Omega^2)}+\frac{\alpha}{4}\left[
\frac{3}{2}(2j+1)y_0^2\left(\frac{F}{M(\omega_{ex}^2-\Omega^2)}\right)^2\right.\nonumber
\\&&\left.+\frac{3}{2}\left(j(j+1)+\frac{1}{2}\right)y_0^4+\frac{3}{8}\left(\frac{F}{M(\omega_{ex}^2-\Omega^2)}\right)^4
\right]-\hbar\omega_{ex}m\nonumber\\&&+\mathcal{O}(\alpha^2).
\end{eqnarray}
In the limit of no driving Eqs. (\ref{GFSstates}) and (\ref{GFSenergy}) yield:
\begin{subequations}
\begin{eqnarray}
E_j&=&\lim_{F\rightarrow 0}\epsilon_{j}=\hbar\Omega\left(j+\frac{1}{2}\right)+\frac{3}{8}\alpha y_0^4 \left(j(j+1)+\frac{1}{2}\right),\label{NOenergies}\\
\ket{j}&=&\lim_{F\rightarrow 0}\ket{\phi_j(t)}=\lim_{F\rightarrow 0}(t\ket{\phi_{j}}\rangle\label{NOstates}\\
&=&\ket{j}_0+\frac{\alpha y_0^4}{4}\left[\frac{\sqrt{j(j-1)}(j-\frac{1}{2})}{2\hbar\Omega}\ket{j-2}_0\right.\nonumber\\
&&\left.+\frac{\sqrt{(j+1)(j+2)}(j+\frac{3}{2})}{-2\hbar\Omega}\ket{j+2}_0\right.\nonumber\\
&&\left.+\frac{\frac{1}{4}\sqrt{j(j-1)(j-2)(j-3)}}{4\hbar\Omega}\ket{j-4}_0+\right.\nonumber\\&&\left.\frac{\frac{1}{4}\sqrt{(j+1)(j+2)(j+3)(j+4)}}{-4\hbar\Omega}\ket{j+4}_0\right]\nonumber,
\end{eqnarray}
\end{subequations}
such that the modifications due to the nonlinearity are exactly those obtained by conventional stationary perturbation theory \cite{Carmen},
where $\{\ket{\ j}_0\}$ are the eigenstates of the undriven harmonic oscillator. Expanding up to second order in the driving amplitude we obtain from Eq. (\ref{GFSstates}) for $\ket{\phi_j(t)}=(t\ket{\phi_{j}}\rangle$ the result:
\begin{eqnarray}
\ket{\phi_j(t)}&=&
\ket{\phi_j(t)}_0+\frac{\alpha}{4}\left[
+\frac{[y_0^4(j+\frac{3}{2})+\frac{3}{2}y_0^2A_\xi^2]\sqrt{(j+1)(j+2)}}{-2\hbar\Omega}\ket{\phi_{j+2}(t)}_0\right.\\&&
+\frac{[y_0^4(j-\frac{1}{2})+\frac{3}{2}y_0^2A_\xi^2]\sqrt{j(j-1)}}{2\hbar\Omega}\ket{\phi_{j-2}(t)}_0\nonumber\\
&&+\frac{\frac{y_0^4}{4}\sqrt{(j+1)(j+2)(j+3)(j+4)}}{-4\hbar\Omega}\ket{\phi_{j+4}(t)}_0\nonumber\\&&
+\frac{\frac{y_0^4}{4}\sqrt{j(j-1)(j-2)(j-3)}}{4\hbar\Omega}\ket{\phi_{j-4}(t)}_0\nonumber\\
&&
+\frac{3}{4}(2j+1)y_0^2A_\xi^2\left[\frac{\exp(-i2\omega_{ex}t)}{\hbar 2\omega_{ex}}
+\frac{\exp(i2\omega_{ex}t)}{-2\hbar \omega_{ex}}\right]\ket{\phi_j(t)}_0\nonumber\\
&&
+\frac{3!\sqrt{2}}{4}(j+1)\sqrt{j+1}A_\xi y_0^3\left[\frac{\exp(-i\omega_{ex}t)}{\hbar \omega_{ex}-\hbar\Omega}
-\frac{\exp(i\omega_{ex}t)}{\hbar \omega_{ex}+\hbar\Omega}\right]\ket{\phi_{j+1}(t)}_0\nonumber
\end{eqnarray}
\begin{eqnarray}
&&
+\frac{3!\sqrt{2}}{4}j\sqrt{j}A_\xi y_0^3\left[\frac{\exp(-i\omega_{ex}t)}{\hbar \omega_{ex}+\hbar\Omega}
+\frac{\exp(i\omega_{ex}t)}{-\hbar \omega_{ex}+\hbar\Omega}\right]\ket{\phi_{j-1}(t)}_0\nonumber\\
&&
+\sqrt{(j+3)(j+2)(j+1)}\frac{2^{3/2}}{4}y_0^3A_\xi\left[\frac{\exp(-i\omega_{ex}t)}{\hbar \omega_{ex}-3\hbar\Omega}
-\frac{\exp(+i\omega_{ex}t)}{\hbar \omega_{ex}+3\hbar\Omega}\right]\ket{\phi_{j+3}(t)}_0\nonumber\\
&&
+\sqrt{j(j-1)(j-2)}\frac{2^{3/2}}{4}y_0^3A_\xi\left[\frac{\exp(-i\omega_{ex}t)}{\hbar \omega_{ex}+3\hbar\Omega}
+\frac{\exp(+i\omega_{ex}t)}{-\hbar \omega_{ex}+3\hbar\Omega}\right]\ket{\phi_{j-3}(t)}_0\nonumber\\
&&
+\frac{3}{4}y_0^2A_\xi^2\sqrt{(j+1)(j+2)}\left[\frac{\exp(-i2\omega_{ex}t)}{\hbar 2\omega_{ex}-2\hbar\Omega}
+\frac{\exp(i2\omega_{ex}t)}{-\hbar 2\omega_{ex}-2\hbar\Omega}\right]\ket{\phi_{j+2}(t)}_0\nonumber\\
&&
+\frac{3}{4}y_0^2A_\xi^2\sqrt{j(j-1)}\left[\frac{\exp(-i2\omega_{ex}t)}{\hbar 2\omega_{ex}+2\hbar\Omega}
+\frac{\exp(i2\omega_{ex}t)}{-\hbar 2\omega_{ex}+2\hbar\Omega}\right]\ket{\phi_{j-2}(t)}_0\left.\right]\nonumber,
\end{eqnarray}
where we used the abbreviation $A_\xi\equiv\frac{F}{M(\omega_{ex}^2-\Omega^2)}$ and $\ket{\phi_j(t)}_0$ are the Floquet states of the linear oscillator.

\section{Perturbative approach for the one-photon resonance}\label{PertApp}
When the nonlinearity becomes a relevant perturbation to the equidistant spectrum of the linear oscillator, it becomes preferable to use the second
 approximation scheme, App II, based on the decomposition Eq. (\ref{App2}).\\
In this case it is convenient to express the Floquet Hamiltonian $\hat{\mathcal{H}}_{\rm DO}$ in the composite Hilbert space $\mathcal{R}\otimes\mathcal{T}$ spanned by the vectors $\ket{j,n}\rangle\equiv\ket{j}\otimes|n)$, where $\ket{j}$ is an eigenstate of the nonlinear oscillator $\hat{H}_{\rm NLO}$ given in Eq. (\ref{NOstates}). Hence, in this basis the Floquet Hamiltonian of the nonlinear oscillator $\hat{\mathcal{H}}_{\rm NLO}$, see Eq. (\ref{matrixeq}) below at vanishing driving amplitude, is diagonal. In contrast, the perturbation $\hat{V}_F=\hat{y}F\cos(\omega_{ex}t)$ is time-dependent and thus non-diagonal also in the Hilbert space $\mathcal{T}$. From the relation:
\begin{eqnarray}
\langle\bra{j,n}\hat{\mathcal{H}}_{\rm DO}\ket{k,n'}\rangle&=&(\hat{H}_{\rm DO})_{jk}^{(n-n')}-\hbar\omega_{ex}n\delta_{jk}\delta_{nn'},
\end{eqnarray}
where $(\hat{H}_{\rm DO})_{jk}^{(n-n')}$ are the Fourier coefficients of the matrix $\bra{j}\hat{H}_{\rm DO}(t)\ket{k}$, it follows
\begin{eqnarray}\label{matrixeq}
\langle\bra{j,n}\hat{\mathcal{H}}_{\rm DO}\ket{k,n'}\rangle&=&E_{j,n}\delta_{kj}\delta_{nn'}+\frac{F}{2}\bra{j}\hat{y}\ket{k}(\delta_{n,n'+1}+\delta_{n,n'-1}),
\end{eqnarray}
with $E_{j,n}=E_{j}-\hbar n\omega_{ex}$ and $E_j$ the energies of the nonlinear oscillator Eq. (\ref{NOenergies}). From Eq. (\ref{matrixeq}) is is thus apparent that two eigenstates $\ket{j,n}\rangle$, $\ket{k,m}\rangle$ of $\hat{\mathcal{H}}_{\rm NLO}$ become degenerate when $E_{j,n}=E_{k,n'}$, i.e. for a driving frequency $\omega_{ex}$ satisfying
\begin{eqnarray}
\hbar\omega_{ex}(n'-n)&=&E_k-E_j.
\end{eqnarray}
Setting $N=n'-n$ one speaks of a $N$-photon resonance. From Eq. (\ref{NOenergies}) for the energies $E_j$ it follows, with $k=j+N$,
\begin{eqnarray}
\hbar\omega_{ex}N&=&E_{j+N}-E_j=N\left[\hbar\Omega+\frac{3}{8}\alpha y_0^4\left(N+1+2j\right)\right].
\end{eqnarray}
In the following we restrict to the one-photon resonance $N=1$, i.e., the quasienergies $E_{j,n}$ and $E_{j+1,n+1}$ are degenerate if $\omega_{ex}=\Omega+\frac{3\alpha y_0^4(j+1)}{4\hbar}\equiv\Omega_j$.\\
Moreover, due to the arbitrariness in the choice of the Brillouin zone index $n$, we fix it in the following to the zeroth Brillouin zone, i.e., $n=0$. For our perturbative treatment we further require that the nonlinearity is large enough that if $E_{j,0}$ is resonant with $E_{j+1,+1}$ the remaining quasi-energy levels 
are off resonance and not involved in the doublet spanned by the two degenerate levels. Having this in mind, we have to restrict ourselves to a certain range of possible values of $\omega_{ex}$, namely to the resonance region such that the chosen doublet remains degenerate or almost degenerate, i.e. for the one-photon resonance: $|\omega_{ex}-\Omega_j|<\frac{3}{4\hbar}\alpha y_0^4$. This results from the fact that if $E_{j,0}=E_{j+1,+1}$, the closest lying levels $E_{j+2,2}$ and $E_{j-1,-1}$ are by $\frac{3}{4\hbar}\alpha y_0^4$ away. Because of the manifold (doublet) structure of the quasi-energy spectrum, we apply in the following
 Van Vleck perturbation theory \cite{Shavitt1980,Cohen1992} and treat the driving as a small perturbation, i.e. $\frac{y_0 F}{2\sqrt{2}}\ll\frac{3}{4\hbar}\alpha y_0^4\ll\hbar\Omega$. Consequently, a consistent treatment in App II requires that either $F^2$ contributions are neglected if we consider the nonlinearity only up to first order, or that both driving and nonlinearity are treated up to second order. As the second order in both parameters is very involved, we restrict to the first order in the nonlinearity and neglect quadratic contributions in the driving strength, as long as their reliability cannot be verified within a different approach, i.e. App I.\\
Within Van Vleck perturbation theory we construct an effective Floquet Hamiltonian 
$\hat{\mathcal{H}}_{{\rm eff}}=\exp(i\hat{S})\hat{\mathcal{H}}_{\rm DO}\exp(-i\hat{S})$ having the same eigenvalues as the original Hamiltonian $\hat{\mathcal{H}}_{\rm DO}$ and not containing matrix elements connecting states belonging to different manifolds. Therefore it is block-diagonal with all quasi-degenerate energy states in one common block. 
To determine the transformation $\hat{S}$ and the effective Hamiltonian $\hat{\mathcal{H}}_{\rm eff}$ we write both as a power series in the driving:
\begin{eqnarray}
\hat{S}&=&\hat{S}^{(0)}+\hat{S}^{(1)}+\hat{S}^{(2)}+\dots\\
\hat{\mathcal{H}}_{\rm eff}&=&\hat{\mathcal{H}}_{\rm eff}^{(0)}+\hat{\mathcal{H}}_{\rm eff}^{(1)}+\hat{\mathcal{H}}_{\rm eff}^{(2)}+\dots.
\end{eqnarray} In \ref{AppVVK} the general formulas for the energies and the states up to second order are provided \cite{Shavitt1980,Cohen1992,Certain,Kirtman}.\\
The zeroth order energies are $E_{j,0}$ and $E_{j+1,+1}$ and the corresponding (quasi)-degenerate Floquet states are:
$|j,0\rangle\rangle$ and $|j+1,+1\rangle\rangle$.\\
The quasi-degenerate block of the effective Hamiltonian in this basis up to second order in the driving strength acquires the form:
\begin{eqnarray}\label{effmatrix}
\hat{\mathcal{H}}_{\rm eff}&=&\left(\begin{array}{cc}
E_{j,0}+E^{(2)--}_j&E^{(1)}_j\\
E^{(1)}_j&E_{j+1,+1}+E^{(2)++}_j\end{array}\right),
\end{eqnarray}
where
\begin{equation}
E^{(1)}_j=\langle\langle j,0|\hat{V}_F|j+1,+1\rangle\rangle=\langle\langle j+1,+1|V_F|j,0\rangle\rangle
=n_1(j)\frac{y_0 F}{2\sqrt{2}},
\end{equation}
and
\begin{eqnarray}\label{secondorder}
E^{(2)--}_j&=&\frac{y_0^2F^2}{8}\left[\frac{n_1^2(j-1)}{E_{j,0}-E_{j-1,-1}}+\frac{n_1^2(j-1)}{E_{j,0}-E_{j-1,+1}}+\frac{n_1^2(j)}{E_{j,0}-E_{j+1,-1}}\right],\\
E^{(2)++}_j&=&\frac{y_0^2F^2}{8}\left[\frac{n_1^2(j+1)}{E_{j+1,+1}-E_{j+2,+2}}+\frac{n_1^2(j+1)}{E_{j+1,+1}-E_{j+2,0}}+\frac{n_1^2(j)}{E_{j+1,+1}-E_{j,2}}\right]\nonumber,
\end{eqnarray}
with
\begin{eqnarray}
n_1(j)&=& \sqrt{j+1}\left[1-\frac{3}{8\hbar\Omega}\alpha(j+1)\right].
\end{eqnarray}
Notice that the unperturbed quasienergies $E_{j,0}$ and $E_{j+1,1}$ are correct up to first order in the nonlinearity.
 For consistency also $n_j^2(j)$ has to be treated up to first order in $\alpha$ only.\\
As shown by Eq. (\ref{EW2}) below it is essential to determine the eigenenergies of $\hat{\mathcal{H}}_{\rm eff}$ up to second order in $F$. They are also the eigenenergies of $\hat{\mathcal{H}}_{\rm DO}$ and read:
\begin{eqnarray}\label{EW2}
\epsilon_{j }^{\mp}&=&\frac{1}{2}\left(E_{j,0}+E_{j+1,+1}+E^{(2)--}_j+E^{(2)++}_j\right)\pm\frac{1}{2}\left[
(E_{j,0}-E_{j+1,+1})^2\right.\\
&&\left.
+2(E_{j,0}-E_{j+1,+1})(E^{(2)--}_j-E^{(2)++}_j)+4E^{(1)^2}_j
\right]^{1/2}\nonumber.
\end{eqnarray}
The convention $\epsilon_{j }^{\mp}$  is chosen such that $\epsilon_{j }^{-}<\epsilon_{j }^{+}$ for $\omega_{ex}<\Omega_j$ , whereas it jumps at resonance such that  $\epsilon_{j }^{-}>\epsilon_{j }^{+}$ for $\omega_{ex}>\Omega_j$.
Because the first order correction in the driving $E_j^{(1)}$ enters Eq. (\ref{EW2}) quadratically, a calculation of the quasienergies up to first order in $F$ merely yields (when $E_{j,0}\neq E_{j+1,1}$) the zeroth order results.
Consequently to be consistent one has to take into account also the second order corrections $E_j^{(2)--}$ and $E_j^{(2)++}$ to the energies.\\
The eigenstates of the block Eq. (\ref{effmatrix}) are determined by:
\begin{eqnarray}
|-_{j,0}\rangle\rangle_{\ \rm eff}&:=&-\sin\frac{\eta_j}{2}|j+1,+1\rangle\rangle+\cos\frac{\eta_j}{2}|j,0\rangle\rangle,\\
|+_{j,1}\rangle\rangle_{\ \rm eff}&:=&\sin\frac{\eta_j}{2}|j,0\rangle\rangle+\cos\frac{\eta_j}{2}|j+1,+1\rangle\rangle\nonumber,
\end{eqnarray}
where
\begin{eqnarray}\label{thetaCarmen}
\tan\eta_j&=&\frac{2|E^{(1)}_j|}{-(E_{j,0}-E_{j+1,+1}+E^{(2)--}_j-E^{(2)++}_j)}.
\end{eqnarray}
In conventional Van Vleck perturbation theory the eigenstates of $\hat{\mathcal{H}}_{\rm DO}$ are obtained by applying a back transformation:
\begin{eqnarray}
\ket{\mp_{j,n}}\rangle&=&\exp(-i\hat{S})\ket{\mp_{j,n}}\rangle_{\rm eff}.
\end{eqnarray}
Expanding the exponential up to first order we obtain for the eigenstates:
\begin{eqnarray}\label{resolv}
\ket{\mp_{j,n}}\rangle&=&\ket{\mp_{j,n}}\rangle_{\rm eff}-i\hat{S}^{(1)}\ket{\mp_{j,n}}\rangle_{ \rm eff},\nonumber\\
&=&\ket{\mp_{j,n}}\rangle_{\rm eff}+\hat{R} \hat{V}_F\ket{\mp_{j,n}}\rangle_{\rm eff}.
\end{eqnarray}
For reasons given in section \ref{Comparison} and the chosen parameter regime of App II we do not determine the second order correction for the states coming from the second order contribution to $\hat{S}$.\\
In the second line of Eq. (\ref{resolv}) we used the fact that we can express the transformation $\hat{S}$ by introducing the reduced resolvent $\hat{R}$, allowing a nice connection to conventional degenerate perturbation theory as shown in \ref{AppVVK}.
From Eq. (\ref{resolv}) it follows:
\begin{eqnarray}\label{statesApp2}
\ket{-_{j,0}}\rangle
&=&-\sin\frac{\eta_j}{2}(1+\hat{R} \hat{V}_F)|j+1,+1\rangle\rangle+\cos\frac{\eta_j}{2}(1+\hat{R} \hat{V}_F)|j,0\rangle\rangle,\\
\ket{+_{j,1}}\rangle
&=&\sin\frac{\eta_j}{2}(1+\hat{R} \hat{V}_F)|j,0\rangle\rangle+\cos\frac{\eta_j}{2}(1+\hat{R} \hat{V}_F)|j+1,+1\rangle\rangle\nonumber,
\end{eqnarray}
where
\begin{eqnarray}\label{VVK1}
\hat{R} \hat{V}_F\ket{j,0}\rangle&=&\sum_{(k,n)\neq\{(j,0),(j+1,+1)\}}\frac{|k,n\rangle\rangle\langle\langle k,n|\hat{V}_F|j,0\rangle\rangle}{E_{j,0}-E_{k,n}}\\
&=&\frac{y_0 F}{2\sqrt{2}}\left(\frac{n_1(j-1)}{E_{j,0}-E_{j-1,-1}}|j-1,-1\rangle\rangle+\frac{n_1(j)}{E_{j,0}-E_{j+1,-1}}|j+1,-1\rangle\rangle\right.\nonumber\\
&&+\frac{n_1(j-1)}{E_{j,0}-E_{j-1,+1}}|j-1,+1\rangle\rangle+\frac{n_3(j,\alpha)}{E_{j,0}-E_{j+3,+1}}|j+3,+1\rangle\rangle\nonumber\\
&&+\frac{n_3(j-3,\alpha)}{E_{j,0}-E_{j-3,+1}}|j-3,+1\rangle\rangle+\frac{n_3(j,\alpha)}{E_{j,0}-E_{j+3,-1}}|j+3,-1\rangle\rangle\nonumber\\
&&\left.+\frac{n_3(j-3,\alpha)}{E_{j,0}-E_{j-3,-1}}|j-3,-1\rangle\rangle\right),\nonumber
\end{eqnarray}
\begin{eqnarray}
\hat{R}\hat{V}_F\ket{j+1,+1}\rangle&=&\sum_{(k,n)\neq\{(j,0),(j+1,+1)\}}\frac{|k,n\rangle\rangle\langle\langle k,n|\hat{V}_F|j+1,+1\rangle\rangle}{E_{j+1,+1}-E_{k,n}}\nonumber\\
&=&\frac{y_0 F}{2\sqrt{2}}\left(\frac{n_1(j)}{E_{j+1,1}-E_{j,2}}|j,+2\rangle\rangle+\frac{n_1(j+1)}{E_{j+1,1}-E_{j+2,2}}|j+2,+2\rangle\rangle\right.\nonumber\\
&&+\frac{n_1(j+1)}{E_{j+1,1}-E_{j+2,0}}|j+2,0\rangle\rangle+\frac{n_3(j-2,\alpha)}{E_{j+1,1}-E_{j-2,0}}|j-2,0\rangle\rangle\nonumber\\
&&+\frac{n_3(j-2,\alpha)}{E_{j+1,1}-E_{j-2,2}}|j-2,+2\rangle\rangle+\frac{n_3(j+1,\alpha)}{E_{j+1,1}-E_{j+4,0}}|j+4,0\rangle\rangle\nonumber\\
&&\left.+\frac{n_3(j+1,\alpha)}{E_{j+1,+1}-E_{j+4,2}}|j+4,+2\rangle\rangle\right),\nonumber
\end{eqnarray}
and
\begin{eqnarray}
n_3(j,\alpha)&=&\frac{\alpha}{16\hbar\Omega}\sqrt{(j+3)(j+2)(j+1)}.\nonumber
\end{eqnarray}
The effect of the transformation is to yield a contribution from states outside the manifold. Notice that in order to obtain the states to first order in $F$ the trigonometric functions $\sin\frac{\eta_j}{2}$ and $\cos\frac{\eta_j}{2}$ should be expanded in powers of $F$.\\
We conclude this section by mentioning that eigenenergies and eigenstates of the Duffing oscillator have been calculated near and at resonance also by Peano et al. \cite{Peano2}. However in \cite{Peano2} the nonlinear undriven Hamiltonian $\hat{H}_{\rm NLO}=\frac{\hat{P}_y^2}{2M}+\frac{M\Omega^2}{2}\hat{y}^2+\frac{\alpha}{4}\hat{y}^4$ is approximated by $\hat{H}_{\rm NLO}\simeq \hbar\Omega\hat{j}+\frac{3}{8}\alpha y_0^4\hat{j}(\hat{j}+1)$, where $\hat{j}$ is the occupation number operator of the undriven linear oscillator. This approximated Hamiltonian is diagonal in the linear oscillator basis and yields the result Eq. (\ref{NOenergies}) for the energies of $\hat{H}_{\rm NLO}$. However, further corrections of order $\alpha$ contained in the eigenstates (\ref{NOstates}) are neglected.
 The results of \cite{Peano2} at finite driving can be retained from Eqs. (\ref{EW2}) and (\ref{statesApp2}) by treating the driving up to first order, by replacing $n_1(j)$ by $\sqrt{j+1}$ and by setting $n_3(j)=0$.\\

\section{Comparison of the outcomes of the two approaches}\label{Comparison}
The approximation scheme in Sec. \ref{GFS}, App I, is valid when the quasienergy spectrum of the linear oscillator is non-degenerate, i.e., away of a $N$-photon resonance. In contrast, the perturbative approach of Sec. \ref{PertApp}, denoted as App II, works at best near a $N$-photon resonance in the quasienergy spectrum of the undriven nonlinear oscillator. Thus a comparison of the outcomes of the two approaches is possible in the frequency regime near resonance, i.e. within $0<|\omega_{ex}-\Omega_j|<\frac{3}{4\hbar}\alpha y_0^4$. Additionallly, as the Van Vleck-based approach is perturbative in the driving, remember $\frac{y_0 F}{2\sqrt{2}}\ll\frac{3}{4\hbar}\alpha y_0^4$, a comparison requires an expansion in $F$ of the results from App I.\\
This section is organized as follows: First the energies and then the matrix elements of the position operator are compared.

\subsection{Comparison of the quasienergies}\label{Compquasien}
We start with the off resonant case $|E_{j,0}-E_{j+1,+1}|=\hbar|\omega_{ex}-\Omega_j|>E_j^{(1)}$ and expand the result in Eq. (\ref{EW2}) up to second order in the driving amplitude $F$:
\begin{eqnarray}
\epsilon_{j}^-&=&E_{j,0}+E^{(2)--}_j+E^{(1)2}_j/(E_{j,0}-E_{j+1,+1}),\\
\epsilon_{j }^{+}&=&E_{j+1,+1}+E^{(2)++}_j-E^{(1)2}_j/(E_{j,0}-E_{j+1,+1})\nonumber.
\end{eqnarray}
Expanding further for consistency the eigenvalues up to first order in the nonlinearity  we obtain:
\begin{eqnarray}
\epsilon_{j }^{-}&=&E_{j,0}+\frac{y_0^2F^2}{8}\left[\frac{2\Omega}{\hbar(\omega_{ex}^2-\Omega^2)}+\frac{3\alpha y_0^4}{\hbar^2}(2j+1)\frac{\Omega^2}{(\omega_{ex}^2-\Omega^2)^2}\right]\\&&+\mathcal{O}(F^3,\alpha^2)\nonumber\\
\epsilon_{j }^{+}&=&E_{j+1,+1}+\frac{y_0^2F^2}{8}\left[\frac{2\Omega}{\hbar(\omega_{ex}^2-\Omega^2)}+\frac{3\alpha y_0^4}{\hbar^2}\frac{(2j+3)\Omega^2}{(\omega_{ex}^2-\Omega^2)^2}\right]+\mathcal{O}(F^3,\alpha^2).\nonumber
\end{eqnarray}
Inserting $y_0=\sqrt{\hbar/(M\Omega)}$, these are exactly the results obtained from App I for $\epsilon_{j,0}$ and $\epsilon_{j+1,1}$ upon expanding Eq. (\ref{GFSenergy}) up to second order in the driving amplitude.
Consequently, as the quasienergies (\ref{GFSenergy}) coincide with the quasienergies from App II away from the resonance, we conclude that Eq. (\ref{EW2}) describes the frequency dependence of the quasienergy up to $\mathcal{O}(F^4)$, over the \textit{whole} parameter regime, i.e., near and far from resonance. Moreover, because the contribution of order $\mathcal{O}(F^4)$ to the quasienergies obtained in App I is state-independent, see Eq. (\ref{GFSenergy}), it drops when differences of quasienergies are considered. In other words the difference of quasienergies coincides in both approaches.

\subsection{Matrix element $y_{lk}(t)$}\label{yalphabeta}
Due to the agreement (in second order in $F$) for the quasienergies and the disagreement for the Floquet states shown in \ref{Compstates}, the question arises whether expectation values of observables also differ in the two approaches. We shall answer this question in the following at the level of the expectation value of the position operator $\hat{y}$.
\subsubsection{$y_{lk}(t)$ in App I}\label{yAppI}
For the linear oscillator the exact result holds:
\begin{eqnarray}
y_{lk}^{(0)}(t)&:=&_0\langle\phi_l(t)|\hat{y}|\phi_k(t)\rangle_0\\
&=&\int dy'dy''\ _0\langle\phi_l(t)|y'\rangle\langle y'|\hat{y}|y''\rangle\langle y''|\phi_k(t)\rangle_0\nonumber\\
&=&\int dy \overline{\phi}_l(y-\xi(t))y\overline{\phi}_k(y-\xi(t))\nonumber\\
&=&\int dy \overline{\phi}_l(y)(y+\xi(t))\overline{\phi}_k(y)\nonumber\\
&=&\frac{y_0}{\sqrt{2}}\left[\sqrt{k+1}\delta_{l,k+1}+\sqrt{k}\delta_{l,k-1}\right]+\xi(t)\delta_{lk}\nonumber.
\end{eqnarray}
where the function $\overline{\phi}_l(y)$ is introduced in Eq. (\ref{flo1b}). Notice that there is \textit{no} second order contribution in the driving to the matrix element $y_{lk}^{(0)}(t)$. This observation will be important later on. 
We now look at the matrix elements of $\hat{y}$ on Floquet states of the driven nonlinear oscillator from App I.
We define for the following:
\begin{eqnarray}
y_{lk}(t)&:=&\bra{\phi_l(t)}\hat{y}\ket{\phi_k(t)}=\sum_n\exp(-in\omega_{ex}t)y_{lk}^{(n)},
\end{eqnarray}
where
\begin{eqnarray}\label{yfourier}
y_{lk}^{(n)}&=&\frac{1}{T_{\omega_{ex}}}\int_0^{T_{\omega_{ex}}}\exp(+in\omega_{ex}t)\bra{\phi_l(t)}\hat{y}\ket{\phi_k(t)}=\langle\bra{\phi_{l,n}}\hat{y}\ket{\phi_{k,0}}\rangle.
\end{eqnarray}
We also define:
\begin{eqnarray}
y_{lk,mn}(t)&:=&\bra{\phi_{l,m}(t)}\hat{y}\ket{\phi_{k,n}(t)}=\exp(-i\omega_{ex}t(n-m))y_{lk}(t).
\end{eqnarray}
We calculate the matrix elements within App I, which is exact in the driving amplitude. In particular from Eq. (\ref{GFSstates}) we obtain:
\begin{eqnarray}
y^{\rm App\ I}_{01,01}(t)
&=&\frac{y_0}{\sqrt{2}}\left[\sqrt{1}+\sqrt{2}\left(c_{20}^{(0)}+c_{20}^{(+2)}\exp(i2\omega_{ex}t)+c_{20}^{(-2)}\exp(-i2\omega_{ex}t)\right)\right.\nonumber\\
&&\left.+\sqrt{1}\left(\exp(-2i\omega_{ex}t)(c_{00}^{(-2)}+c_{11}^{(2)})+\exp(2i\omega_{ex}t)(c_{00}^{(2)}+c_{11}^{(-2)})\right)\right]\cdot\nonumber\\&&\exp(-i\omega_{ex}t)\nonumber,\\
y^{\rm App\ I}_{00,00}(t)&=&\xi(t)+\frac{y_0}{\sqrt{2}}\left[2\cos(\omega_{ex}t)(c_{10}^{(1)}+c_{10}^{(-1)})+2\cos(3\omega_{ex}t)(c_{10}^{(3)}+c_{10}^{(-3)})\right]\nonumber,\\
y^{\rm App\ I}_{10,10}(t)&=&\left(y^{\rm App\ I}_{01,01}(t)\right)^*,\nonumber\\
y^{\rm App\ I}_{11,11}(t)&=&\xi(t)+\frac{y_0}{\sqrt{2}}\left[2\cos(\omega_{ex}t)(c_{01}^{(1)}+c_{01}^{(-1)})+2\cos(3\omega_{ex}t)(c_{01}^{(3)}+c_{01}^{(-3)})\right.\nonumber\\&&\left.+2\sqrt{2}\cos(\omega_{ex}t)(c_{21}^{(1)}+c_{21}^{(-1)})+2\sqrt{2}\cos(3\omega_{ex}t)(c_{21}^{(3)}+c_{21}^{(-3)})
\right]\nonumber.
\end{eqnarray}
In the last derivations we used the coefficients $c_{jk}^{(n)}$ introduced in Eq. (\ref{ccoeff}) 
and the symmetry relations:
\begin{eqnarray}
c_{jj}^{(\pm4)}&=&c_{kk}^{(\pm4)},\quad j\neq k\\
c_{jk}^{(n)}&=&-c_{kj}^{(-n)}.\nonumber
\end{eqnarray}
Inserting the actual form of the coefficients $c_{jk}^{(n)}$ we obtain:
\begin{eqnarray}
y^{\rm App\ I}_{00,00}&=&\frac{F}{M(\omega_{ex}^2-\Omega^2)}\cos(\omega_{ex}t)\left[1+\frac{3F\alpha y_0^2 }{2 M(\omega_{ex}^2-\Omega^2)}\right]\\
&&+\mathcal{O}(\alpha F^2),\nonumber\\
y^{\rm App\ I}_{10,10}&=&\frac{y_0}{\sqrt{2}}\sqrt{1}\left(1-\frac{3\alpha y_0^4}{8\hbar\Omega}\right)\exp(+i\omega_{ex}t)+\mathcal{O}(\alpha F^2),\\
y^{\rm App\ I}_{11,11}&=&\frac{F}{M(\omega_{ex}^2-\Omega^2)}\cos(\omega_{ex}t)\left[1+\frac{9 F\alpha y_0^2 }{2 M (\omega_{ex}^2-\Omega^2)}\right]\\&&+\mathcal{O}(\alpha F^2).\nonumber
\end{eqnarray}
We observe that, as it is well known from the driven linear oscillator, terms of order $F^2$ or higher are absent at zero nonlinearity.
The nonlinearity introduces a correction of $\mathcal{O}(\alpha F^2)$, which we neglect in the following. This is the reason why we neglected in Eq. (\ref{resolv}) the second order contribution in the back transformation, as we can not rely on the $F^2$ contributions of the eigenstates originating from App II (see discussion in section \ref{PertApp} concerning the limitations of App II and \ref{Compstates}), as long as these do not coincide with App I.\\

\subsubsection{$y_{lk}(t)$ in the perturbative approach App II}
We calculate $y_{lk}(t)$ for the one-photon resonance for the case $j=0$. In the following we define $\ket{\phi_{0,0}(t)}\equiv(t|-_{0,0}\rangle\rangle$ and $\ket{\phi_{1,1}(t)}\equiv(t|+_{0,1}\rangle\rangle$. Then
\begin{eqnarray}
y_{00,00}^{\rm App\ II}(t)&\equiv&\langle\phi_{0,0}(t)|\hat{y}|\phi_{0,0}(t)\rangle\nonumber\\
&=&\sum_n\exp(-in\omega_{ex}t)y_{00}^{(n)}\nonumber\\
&=&\sum_n\exp(-in\omega_{ex}t)\langle\bra{-_{0,n}}\hat{y}\ket{-_{0,0}}\rangle.
\end{eqnarray}
To proceed we use Eqs. (\ref{statesApp2}) and (\ref{VVK1}) for $j=0$. Using the relation $\hat{y}=\frac{y_0}{\sqrt{2}}(\hat{a}+\hat{a}^\dagger)$, where $\hat{a}$ and $\hat{a}^\dagger$ are the annihilation and creation operators of the linear oscillator, we calculate the matrix elements:
\begin{eqnarray}
\langle \bra{0,n}(1+\hat{R}\hat{V}_F)^\dagger(\hat{a}+\hat{a}^\dagger)(1+\hat{R}\hat{V}_F)\ket{0,0}\rangle
&=&\frac{y_0 F}{2\sqrt{2}}\frac{n_1^2(0)}{E_{0,0}-E_{1,-1}}(\delta_{n,1}+\delta_{n,-1})\nonumber\\ 
&\equiv& A_{--}(F)(\delta_{n,1}+\delta_{n,-1}),\nonumber
\end{eqnarray}
\begin{eqnarray}
\langle\bra{1,n+1}(1+\hat{R}\hat{V}_F)^\dagger(\hat{a}+\hat{a}^\dagger)(1+\hat{R}\hat{V}_F)|\ket{1,1}
&=&\frac{y_0 F}{2\sqrt{2}}\left[\frac{n_1^2(0)}{E_{1,1}-E_{0,2}}\right.\nonumber\\
\left.+\frac{n_1^2(1)}{E_{1,1}-E_{2,2}}+\frac{n_1^2(1)}{E_{1,1}-E_{2,0}}\right](\delta_{n,1}+\delta_{n,-1}))\nonumber
&\equiv& A_{++}(F)(\delta_{n,1}+\delta_{n,-1}),\nonumber
\end{eqnarray}
\begin{eqnarray}
\langle\bra{1,n+1}(1+\hat{R}\hat{V}_F)^\dagger(\hat{a}+\hat{a}^\dagger)(1+\hat{R}\hat{V}_F)\ket{0,0}\rangle&=&n_1(0)\delta_{n,-1}\equiv A_{+-}\delta_{n,-1},\nonumber\\
\langle \bra{0,n}(1+\hat{R}\hat{V}_F)^\dagger(\hat{a}+\hat{a}^\dagger)(1+\hat{R}\hat{V}_F)\ket{1,1}\rangle&=&n_1(0)\delta_{n,1}\equiv A_{-+}\delta_{n,1}\nonumber.\\
\end{eqnarray}
Note that $A_{+-}=A_{-+}=n_1(0)$ is independent of the driving. Consequently we find the result:
\begin{eqnarray}\label{y00}
y_{00,00}^{\rm App\ II}(t)&=&y_{00}^{(+1)}\exp(-i\omega_{ex}t)+y_{00}^{(-1)}\exp(i\omega_{ex}t),
\end{eqnarray}
with
\begin{eqnarray}
y_{00}^{(+1)}&=&y_{00}^{(-1)}\\
&\equiv&\frac{y_0}{\sqrt{2}}\left(\sin^2\frac{\eta_0}{2}A_{++}(F)-\sin\frac{\eta_0}{2}\cos\frac{\eta_0}{2}A_{+-}+
\cos^2\frac{\eta_0}{2}A_{--}(F)\right).\nonumber
\end{eqnarray}
The other matrix elements are obtained in the same way. We give only the results:
\begin{eqnarray}
y_{11,11}^{\rm App\ II}(t)&=&\bra{\phi_{1,1}(t)}\hat{y}\ket{\phi_{1,1}(t)}=\sum_n\exp(-in\omega_{ex}t)\langle\bra{+_{0,n+1}}\hat{y}\ket{+_{0,1}}\rangle\\
&=&y_{11}^{(+1)}\exp(-i\omega_{ex}t)+y_{11}^{(-1)}\exp(i\omega_{ex}t)\nonumber
\end{eqnarray}
with
\begin{eqnarray}
y_{11}^{(+1)}&=&y_{11}^{(-1)}\\
&\equiv&\frac{y_0}{\sqrt{2}}\left(\sin^2\frac{\eta_0}{2}A_{--}(F)+\sin\frac{\eta_0}{2}\cos\frac{\eta_0}{2}A_{+-}+\cos^2\frac{\eta_0}{2}A_{++}(F)\right),\nonumber
\end{eqnarray}
\begin{eqnarray}
y_{10,10}^{\rm App\ II}(t)&=&\bra{\phi_{1,1}(t)}\hat{y}\ket{\phi_{0,0}(t)}=\sum_n\exp(-in\omega_{ex}t)\langle\bra{+_{0,n+1}}\hat{y}\ket{-_{0,0}}\rangle\\
&=&y_{10}^{(+1)}\exp(-i\omega_{ex}t)+y_{10}^{(-1)}\exp(i\omega_{ex}t)\nonumber
\end{eqnarray}
with
\begin{eqnarray}
y_{10}^{(+1)}&\equiv&\frac{y_0}{\sqrt{2}}\left(-\sin^2\frac{\eta_0}{2}A_{-+}+\sin\frac{\eta_0}{2}\cos\frac{\eta_0}{2}[A_{--}(F)-A_{++}(F)]\right),\\
y_{10}^{(-1)}&\equiv&\frac{y_0}{\sqrt{2}}\left(+\sin\frac{\eta_0}{2}\cos\frac{\eta_0}{2}[A_{--}(F)-A_{++}(F)]+
\cos^2\frac{\eta_0}{2}A_{+-}\right),\nonumber
\end{eqnarray}
and
\begin{eqnarray}
y_{01,01}^{\rm App\ II}(t)&=&\bra{\phi_{0,0}(t)}\hat{y}\ket{\phi_{1,1}(t)}=\sum_n\exp(-in\omega_{ex}t)\langle\bra{-_{0,n}}\hat{y}\ket{+_{0,1}}\rangle\\
&=&y_{01}^{(+1)}\exp(-i\omega_{ex}t)+y_{01}^{(-1)}\exp(i\omega_{ex}t)\nonumber
\end{eqnarray}
with
\begin{eqnarray}\label{y011}
y_{01}^{(+1)}=y_{10}^{(-1)},\\
y_{01}^{(-1)}=y_{10}^{(+1)}\nonumber.
\end{eqnarray}
Hence, $y^{\rm App\ II}_{10,10}(t)=\left(y^{\rm App\ II}_{01,01}(t)\right)^*$.\\
To compare we now expand the matrix elements in the driving strength and in the nonlinearity up to first order, using:
\begin{eqnarray}
\sin\frac{\eta_j}{2}&=&\frac{y_0 Fn_1(j)}{2\sqrt{2}(E_{j+1,+1}-E_{j,0})}+\mathcal{O}(F^3,\alpha^2),\\
\cos\frac{\eta_j}{2}&=&1-\frac{1}{16}\frac{y_0^2F^2n_1^2(j)}{(E_{j+1,+1}-E_{j,0})^2}+\mathcal{O}(F^4,\alpha^2).
\end{eqnarray}
Consequently Eqs. (\ref{y00})- (\ref{y011}) yield:
\begin{eqnarray}
y^{\rm App\ II}_{00,00}(t)&=&\frac{F}{M(\omega_{ex}^2-\Omega^2)}\cos(\omega_{ex}t)\left[1+\frac{3 F\alpha y_0^2 }{2 M (\omega_{ex}^2-\Omega^2)}\right]\\&&+\mathcal{O}(\alpha^2, F^2),\nonumber\\
y^{\rm App\ II}_{10,10}(t)&=&\frac{y_0}{\sqrt{2}}\left(1-\frac{3\alpha y_0^4}{8\hbar\Omega}\right)\exp(i\omega_{ex}t)+\mathcal{O}(\alpha^2, F^2),\\
y^{\rm App\ II}_{11,11}(t)&=&\frac{F}{M(\omega_{ex}^2-\Omega^2)}\cos(\omega_{ex}t)\left[1+\frac{9 F\alpha y_0^2 }{2 M (\omega_{ex}^2-\Omega^2)}\right]\\&&+\mathcal{O}(\alpha^2, F^2),\nonumber
\end{eqnarray}
Hence App I reproduces the expressions for the matrix elements $y_{lk}(t)$ obtained in App II up to first order in the driving $F$ near the one-photon resonance. Note that, as in App I, the difference between $y^{\rm App\ II}_{00,00}(t)$ and $y^{\rm App\ II}_{11,11}(t)$ in the nonlinear contribution arises due to the contribution of states neighbouring the (quasi)-degenerate states. App I is not valid at resonance, as the generated degeneracy is not included in the perturbative treatment. For App II the resonance condition is essential for generating a doublet of degenerate levels, requiring a finite nonlinearity. However the resulting energies and matrix elements can be expanded in the off-resonant regime and coincide with the results obtained within App I. Combining both approaches we can describe the whole range of possible driving frequencies, using Eqs. (\ref{y00})-(\ref{y011}).

\section{Dissipative dynamics}\label{Dissipation}
To include dissipative effects we use the system-bath approach introduced by Caldeira and Leggett \cite{CaldeiraLeggett}, where the bath is composed by $\mathcal{N}$ harmonic oscillators with coordinate $\hat{x}_j$ momentum $\hat{p}_j$ and frequency $\omega_j$ and is coupled bilinearly to the system degrees of freedom. The interaction between system and bath is encapsulated in the coupling constants $c_j$. We have also introduced a counter term, proportional to $\hat{y}^2$, which accounts for the renormalization of the potential of the system due to the coupling.\\
The total Hamiltonian of the system plus bath is given by:
\begin{eqnarray}
\hat{H}(t)&=&\hat{H}_{\rm DO}(t)+\hat{H}_{\rm B}+\hat{H}_{\rm DO+B},
\end{eqnarray}
where
\begin{eqnarray}
\hat{H}_{\rm B}&=&\sum_{j=1}^\mathcal{N}\left[\frac{\hat{p}_j^2}{2m_j}+\frac{1}{2}m_j\omega_j^2 \hat{x}_j^2\right]\nonumber\\
\hat{H}_{\rm DO+B}&=&-\hat{y}\sum_{j=1}^\mathcal{N} c_j \hat{x}_j+\hat{y}^2\sum_{j=1}^\mathcal{N}\frac{c_j^2}{2m_j\omega_j^2} \nonumber.
\end{eqnarray}
In order to specifiy the effect of the bath it is convenient to introduce the spectral density of the bath:
\begin{eqnarray}
J(\omega)&=&\frac{\pi}{2}\sum_{j=1}^\mathcal{N}\frac{c_j^2}{m_j\omega_j}\delta(\omega-\omega_j). 
\end{eqnarray}
For a large number of bath oscillators the spectral density will become a smooth function and in the following calculations we use the Ohmic case: $J(\omega)=M\gamma\omega,$ leading to memoryless friction and white noise in the classical limit. We will assume for further calculation that the Duffing oscillator (DO) and bath are uncorrelated at time $t=0$:
\begin{eqnarray}
\hat{\rho}_{\rm DO+B}(0)&=&\hat{\rho}_{\rm DO}(0)\otimes\frac{\exp(-\hat{H}_{\rm B}/k_BT)}{{\rm{tr_B}}\exp(-\hat{H}_{\rm B}/k_BT)},
\end{eqnarray}
where $\hat{\rho}_{\rm DO}(0)$ is the density operator of the Duffing oscillator at time $t=0$. Because the bath consists of infinite degrees of freedom we assume the effects of the interaction with the DO system on the bath to dissipate away quickly, such that the bath remains in thermal equilibrium for all times $t$.\\
We wish to obtain an equation of motion for the reduced density operator $\hat{\rho}_{\rm DO}(t)={\rm{tr_B}}\hat{\rho}_{\rm DO+B}(t)$.
Following \cite{Louisell,Blümel1,Blümel2,Blum,Kohler1,Kohler2} a Born-Markov approximation is applied and a Floquet-Markov master equation for 
the reduced density operator expressed in the Floquet basis of the DO is derived:
\begin{eqnarray}\label{q5}
\dot{\rho}_{\alpha\beta}(t)&=&-\frac{i}{\hbar}(\epsilon_\alpha-\epsilon_\beta)\rho_{\alpha\beta}(t)\\
&&+\sum_{\alpha'\beta'}\sum_{nn'}\exp[-i(n+n')\omega_{ex}t][(N_{\alpha\alpha',-n}+N_{\beta\beta',n'})y_{\alpha\alpha'}^{(n)}y_{\beta'\beta}^{(n')}\rho_{\alpha'\beta'}\nonumber\\
&&-N_{\beta'\alpha',-n}y_{\alpha\beta'}^{(n')}y_{\beta'\alpha'}^{(n)}\rho_{\alpha'\beta}-N_{\alpha'\beta',n'}y_{\beta'\alpha'}^{(n')}y_{\alpha'\beta}^{(n)}\rho_{\alpha\beta'}\nonumber],
\end{eqnarray}
where  $\rho_{\alpha\beta}(t)=\langle\phi_{\alpha}(t)|\hat{\rho}_{\rm DO}(t)|\phi_\beta(t)\rangle$. A Lamb-shift contribution  was disregarded. The other quantities entering Eq. (\ref{q5}) are
\begin{eqnarray}\label{q14}
N_{\alpha\beta,n}&=&N(\epsilon_\alpha-\epsilon_\beta+\hbar n\omega_{ex}),\\
N(\epsilon)&=&\frac{J(|\epsilon|)}{\hbar}[n_{th}(|\epsilon|)+\theta(-\epsilon)]\nonumber,
\end{eqnarray}
where $\theta(x)$ is the Heaviside function and $n_{th}$ is the bosonic thermal occupation number $n_{th}(\epsilon)=\frac{1}{2}\left[\coth\left(\frac{\epsilon}{2k_BT}\right)-1\right]$. 
Additionally, $y_{\alpha\beta}^{(n)}$ are the Fourier coefficients defined of the matrix elements calculated in Sec. \ref{yAppI} , see Eq. (\ref{yfourier}).
For additional simplification we perform a moderate rotating-wave approximation consisting in averaging the time-dependent terms in the bath part over the driving period $T_{\omega_{ex}}=2\pi/\omega_{ex}$ \cite{Kohler1,Kohler2}:
\begin{eqnarray}\label{q7}
\dot{\overline{\rho}}_{\alpha\beta}(t)&=&-\frac{i}{\hbar}(\epsilon_\alpha-\epsilon_\beta)\overline{\rho}_{\alpha\beta}(t)+\sum_{\alpha'\beta'}\mathcal{L}_{\alpha\beta,\alpha'\beta'}\overline{\rho}_{\alpha'\beta'},
\end{eqnarray}
where $\overline{\rho}$ indicates the time average and the dissipative transition rates are:
\begin{eqnarray}
\mathcal{L}_{\alpha\beta,\alpha'\beta'}&=&\sum_n(N_{\alpha\alpha',-n}+N_{\beta\beta',-n})y_{\alpha\alpha'}^{(n)}y_{\beta'\beta}^{(-n)}\\
&&-\delta_{\alpha\alpha'}\sum_{\alpha'',n}N_{\alpha''\beta',-n}y_{\beta'\alpha''}^{(-n)}y_{\alpha''\beta}^{(n)}
-\delta_{\beta\beta'}\sum_{\beta'',n}N_{\beta''\alpha',-n}y_{\alpha\beta''}^{(-n)}y_{\beta''\alpha'}^{(n)}\nonumber.
\end{eqnarray}
The reason for its application is that it yields a time-independent stationary solution $\rho_{\alpha\beta}^{\rm st}=\lim_{t\rightarrow\infty}\overline{\rho}_{\alpha\beta}(t)$, which can be obtained by solving the linear system of equations:
\begin{eqnarray}\label{q4}
0&=&-\frac{i}{\hbar}(\epsilon_\alpha-\epsilon_\beta)\rho_{\alpha\beta}^{\rm st}+\sum_{\alpha'\beta'}\mathcal{L}_{\alpha\beta,\alpha'\beta'}\rho_{\alpha'\beta'}^{\rm st}.
\end{eqnarray}

\section{Observable for the nonlinear response}\label{Observable}
Using Eq. (\ref{q4}) corresponds to describe the long time limit, where the system has already reached the steady state. We calculate the mean value of the position operator in the stationary state, the nonlinear response:
\begin{eqnarray}
\langle \hat{y}\rangle_{\rm st}&=&\lim_{t\rightarrow\infty}{\rm tr}\{\hat{\overline{\rho}}_{\rm DO}(t)\hat{y}\}=\sum_{\alpha\beta}\rho_{\alpha\beta}^{\rm st}y_{\beta\alpha}(t),
\end{eqnarray}
where ${\rm tr}\{\}$ denotes the trace over the oscillator degrees of freedom. Upon focussing on driving frequencies near and at the first resonance, the solution in the long time limit for a driven system is given by:
\begin{eqnarray}
\langle \hat{y}\rangle_{\rm st}\simeq A\cos(\omega_{ex}t+\phi),
\end{eqnarray}
where higher harmonics have been neglected. We introduced the amplitude: 
\begin{eqnarray}\label{q20}
A&=&2|\sum_{\alpha\beta}\rho_{\alpha\beta}^{\rm st}y_{\beta\alpha}^{(+1)}|,
\end{eqnarray}
and phase shift 
\begin{eqnarray}
\phi&=&+\pi\theta\left(-\mbox{Re}\sum_{\alpha\beta}\rho_{\alpha\beta}^{\rm st}y_{\beta\alpha}^{(+1)}\right)-\arctan\left[\frac{\mbox{Im}\sum_{\alpha\beta}\rho_{\alpha\beta}^{\rm st}y_{\beta\alpha}^{(+1)}}{\mbox{Re}\sum_{\alpha\beta}\rho_{\alpha\beta}^{\rm st}y_{\beta\alpha}^{(+1)}}\right].
\end{eqnarray} 

\subsection{One-photon resonance using PSA}
We solve the master equation close to resonance in the low temperature regime, imposing a partial secular approximation (PSA) which amounts to take only coherences of the involved resonant levels into account. 
For an intermediate damping strength, i.e. the damping is of the order of the splitting of the one-photon resonance or larger, the system of equations is:
\begin{eqnarray}
0&=&\mathcal{L}_{00,00}\rho_{00}(t)+\mathcal{L}_{00,11}\rho_{11}(t)+2\mathcal{L}_{00,01}\mbox{Re}\left[\rho_{01}(t)\right],\\
0&=&-i(\epsilon_0-\epsilon_1)\rho_{01}+\mathcal{L}_{01,00}\rho_{00}+\mathcal{L}_{01,11}\rho_{11}+\mathcal{L}_{01,01}\rho_{01}+\mathcal{L}_{01,10}\rho_{01}^*\nonumber,\\
\rho_{11}&=&1-\rho_{00}\nonumber,
\end{eqnarray}
where for simplicity we omit the superscript 'st'.\\
If the damping $\gamma$ is smaller than the smallest splitting we can even neglect the coherences completely:
\begin{eqnarray}\label{lowdamp}
0&=& \sum_\beta\mathcal{L}_{\alpha\alpha,\beta\beta}\rho_{\beta\beta}.
\end{eqnarray}
Note that we use as shorthand abbreviations: $\epsilon_0:=\epsilon_{0}^-$ with eigenstate $\ket{0}=\ket{\phi_{0,0}(t)}$ and $\epsilon_1:=\epsilon_{0}^+$  with $\ket{1}=\ket{\phi_{1,1}(t)}$, respectively. Solving the system of equation for the stationary solution, we obtain:
\begin{eqnarray}\label{eqsys}
\rho_{00}&=&\left\{-\mathcal{L}_{00,11}[\mathcal{L}_{01,01}^2-\mathcal{L}_{01,10}^2+(\epsilon_0-\epsilon_1)^2]+2\mathcal{L}_{00,01}\mathcal{L}_{01,11}(\mathcal{L}_{01,01}\right.\nonumber\\&&\left.-\mathcal{L}_{01,10})\right\}/\left\{(\mathcal{L}_{00,00}-\mathcal{L}_{00,11})[\mathcal{L}_{01,01}^2-\mathcal{L}_{01,10}^2+(\epsilon_0-\epsilon_1)^2]\right.\nonumber\\
&&\left.-2\mathcal{L}_{00,01}(\mathcal{L}_{01,00}-\mathcal{L}_{01,11})(\mathcal{L}_{01,01}-\mathcal{L}_{01,10})\right\} \nonumber,\\
\mbox{Re}\rho_{01}&=&\frac{-(\mathcal{L}_{01,01}-\mathcal{L}_{01,10})[\mathcal{L}_{01,11}+(\mathcal{L}_{01,00}-\mathcal{L}_{01,11})\rho_{00}]}{\mathcal{L}_{01,01}^2-\mathcal{L}_{01,10}^2+(\epsilon_0-\epsilon_1)^2}\nonumber,\\
\mbox{Im}\rho_{01}&=&\frac{\epsilon_0-\epsilon_1}{(\mathcal{L}_{01,01}-\mathcal{L}_{01,10})}\mbox{Re}\rho_{01}.
\end{eqnarray}
To simplify the rates and obtain analytic results we restrict to low temperatures, $k_BT\ll\hbar\omega_{ex}$. Moreover, we consider the vicinity of a resonance, such that $|\epsilon_\alpha-\epsilon_\beta|$ is of the order of the minimal splitting, proportional to $F$, and within the limit of validity of App II, we obtain $|\epsilon_\alpha-\epsilon_\beta|\ll\hbar\omega_{ex}$. This corresponds to consider only spontaneous emission (see Eq. (\ref{q14})): $N_{\alpha\beta,-1}\simeq J(|\omega_{ex}|)$ and $N_{\alpha\beta,1}\simeq 0$.\\ So we obtain:
\begin{eqnarray}
\mathcal{L}_{\alpha\beta,\alpha'\beta'}&=&J(|\omega_{ex}|)\left[2y_{\alpha\alpha'}^{(1)}y_{\beta'\beta}^{(-1)}-\delta_{\alpha\alpha'}\sum_{\alpha''}y_{\beta'\alpha''}^{(-1)}y_{\alpha'',\beta}^{(1)}\right.\\
&&\left.
-\delta_{\beta\beta'}\sum_{\beta''}y_{\alpha\beta''}^{(-1)}y_{\beta''\alpha'}^{(1)}\right]\nonumber,
\end{eqnarray}
with $J(|\omega_{ex}|)=M\gamma|\omega_{ex}|.$

\subsection{One-photon resonance versus antiresonance}
We use these approximate low temperature rates and solve the master equation near the one-photon resonance to calculate the amplitude and phase of the steady state. We calculate the amplitude $A$ in Eq. (\ref{q20}) for the one-photon resonance using the energies Eq. (\ref{EW2}) up to second order in the driving and first in the nonlinearity and the expectation values Eqs. (\ref{y00})-(\ref{y011}) up to both first order in the driving and in the nonlinearity.
\subsubsection{The amplitude in lowest order of the damping}
\begin{figure}[h!]
\begin{center}
\includegraphics[width=12.0cm]{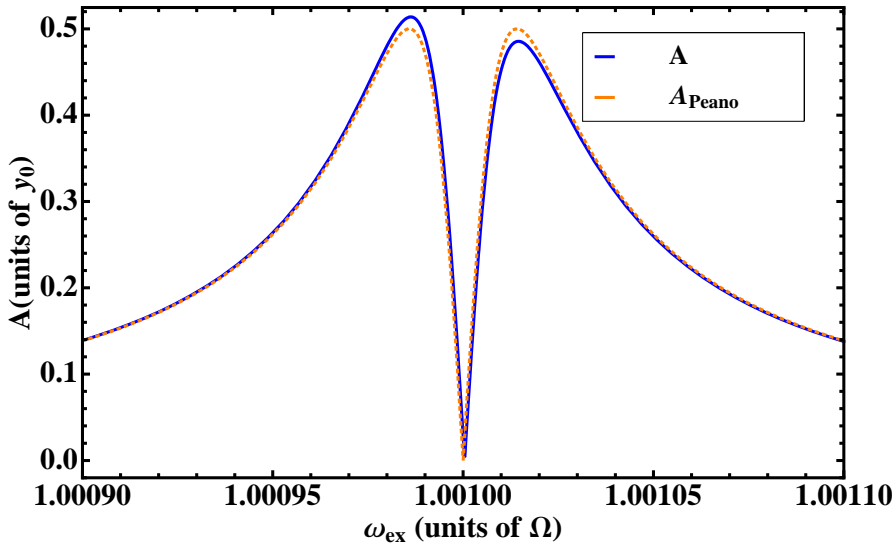}
\caption{Amplitude $A$ at low damping ($\gamma\ll y_0Fn_1(0)/\sqrt{2}$) versus the driving frequency $\omega_{ex}$.
Chosen parameters are: $\Omega=1$,
 $y_0^4\alpha/(\hbar\Omega)=4/3\cdot 10^{-3}$, $y_0F/(\hbar\Omega)=2^{3/2}\cdot10^{-5}$, $k_b T/(\hbar\Omega)=10^{-3}$ and $\gamma/\Omega=10^{-7}$. For comparison we give also the result obtained by Peano et al. in Eq. (43) in \cite{Peano2}.\label{Alowg}}
\end{center}
\end{figure}

First we start with the analytical result for very low damping, where we only have to solve Eq. (\ref{lowdamp}) resulting in:
\begin{eqnarray}
\rho_{00}&=&\frac{-\mathcal{L}_{00,11}}{(\mathcal{L}_{00,00}-\mathcal{L}_{00,11})}=\frac{y_{01,1}^2}{(y_{10,1}^2+y_{01,1}^2)}+\mathcal{O}(\gamma)\\
\rho_{11}&=&1-\rho_{00}+\mathcal{O}(\gamma)\nonumber.
\end{eqnarray}
Therefore the amplitude Eq. (\ref{q20}) in lowest order, i.e. zeroth order, in the damping is determined to be:\\
\begin{eqnarray}\label{Ampl}
A&=&
2\left|\left[y_0 \left(2 \sin \eta_04 A_{+-} \cos \eta_0 A_{--}^2(F)\right.\right.\right.\\&&
+\left((5 \cos (2 \eta_0 )+3) A_{+-}^2-4 A_{++}(F) \sin (2 \eta_0 ) A_{+-}\right)A_{--}(F)\nonumber\\&&\left.\left.
+6A_{+-}^2 A_{++}(F) \sin ^2\eta_0-2 A_{+-}^3 \sin(2 \eta_0 )\right)\right]\nonumber\\&&
/\left[
2 \sqrt{2} \left((\cos (2 \eta_0 )+3)A_{+-}^2+2 (A_{--}(F)-A_{++}(F)) \sin (2 \eta_0 )A_{+-}\right.\right.\nonumber\\&&\left.\left.\left.+2 (A_{--}(F)-A_{++}(F))^2 \sin ^2(\eta_0
   )\right)\right]\right|\nonumber.
\end{eqnarray}
The actual form for the amplitude is shown in the Figure \ref{Alowg}. Interestingly, an antiresonance occurs, as already predicted in \cite{Fistul,Peano2}.\\
For comparison we show the result obtained by Peano et al. \cite{Peano2}. We see that both curves show an exact antiresonance, but the dip position is slightly different and our result shows an asymmetric line shape. The reason for the differences are explained in the following. The condition for an exact antiresonance, $A=0$, is $\rho_{00}y_{00}^{(1)}=-\rho_{11}y_{11}^{(1)}$.
As the result of \cite{Peano2} is given in lowest order without back transformation, the dip position occurs at the one-photon resonance, when the driving is such that $\sin\frac{\eta_0}{2}=\cos\frac{\eta_0}{2}=1/\sqrt{2}$. This corresponds to resonance in \cite{Peano2} due to $\rho_{00}=\rho_{11}=\frac{1}{2}$ and with $y_{00}^{(+1)}=-y_{11}^{(1)}=-\sin\frac{\eta_0}{2}\cos\frac{\eta_0}{2}\frac{y_0}{\sqrt{2}}$, yielding a symmetric shape of the amplitude, 
seen in Figure \ref{Alowg}. When all contributions linear in $\alpha$ are included, we find at resonance:
\begin{equation}
\rho_{00}=\frac{(A_{+-}+A_{--}(F)-A_{++}(F))^2}{2\left[A_{+-}^2+(A_{--}(F)-A_{++}(F))^2\right]}\neq\frac{1}{2},
\end{equation}
and $y_{00}^{(+1)}\neq-y_{11}^{(1)}$. Moreover $A_{\eta_0=\pi/4}\neq0$. 
The reason for not obtaining an equal population of the involved levels at resonance is due to the back transformation leading to the dressing of the eigenstates by states outside the quasi-degenerate doublet. In general, the dip position is determined by minimizing the amplitude $A$, Eq. (\ref{Ampl}), with respect to the external driving frequency. As the driving enters in both $A_{++}$, $A_{--}$ and in $\eta_0$, the amplitude acquires a nontrivial $\omega_{ex}$-dependence, such that the minimization can only be done numerically.
The antiresonance does not occur exactly at $\eta_0=\pi/4$, but 
it is shifted to little larger values (deviation $\propto 10^{(-7)}$) with respect to the resonant case. This deviation arises due to the interplay of terms involving trigonometric functions of $\eta_0$ and explicitly driving-dependent prefactors, $A_{++}(F)$ and $A_{--}(F)$, resulting from the back transformation.
\subsubsection{Solution for the amplitude including higher orders in the damping}
\begin{figure}[h!]
\includegraphics[width=12.0cm]{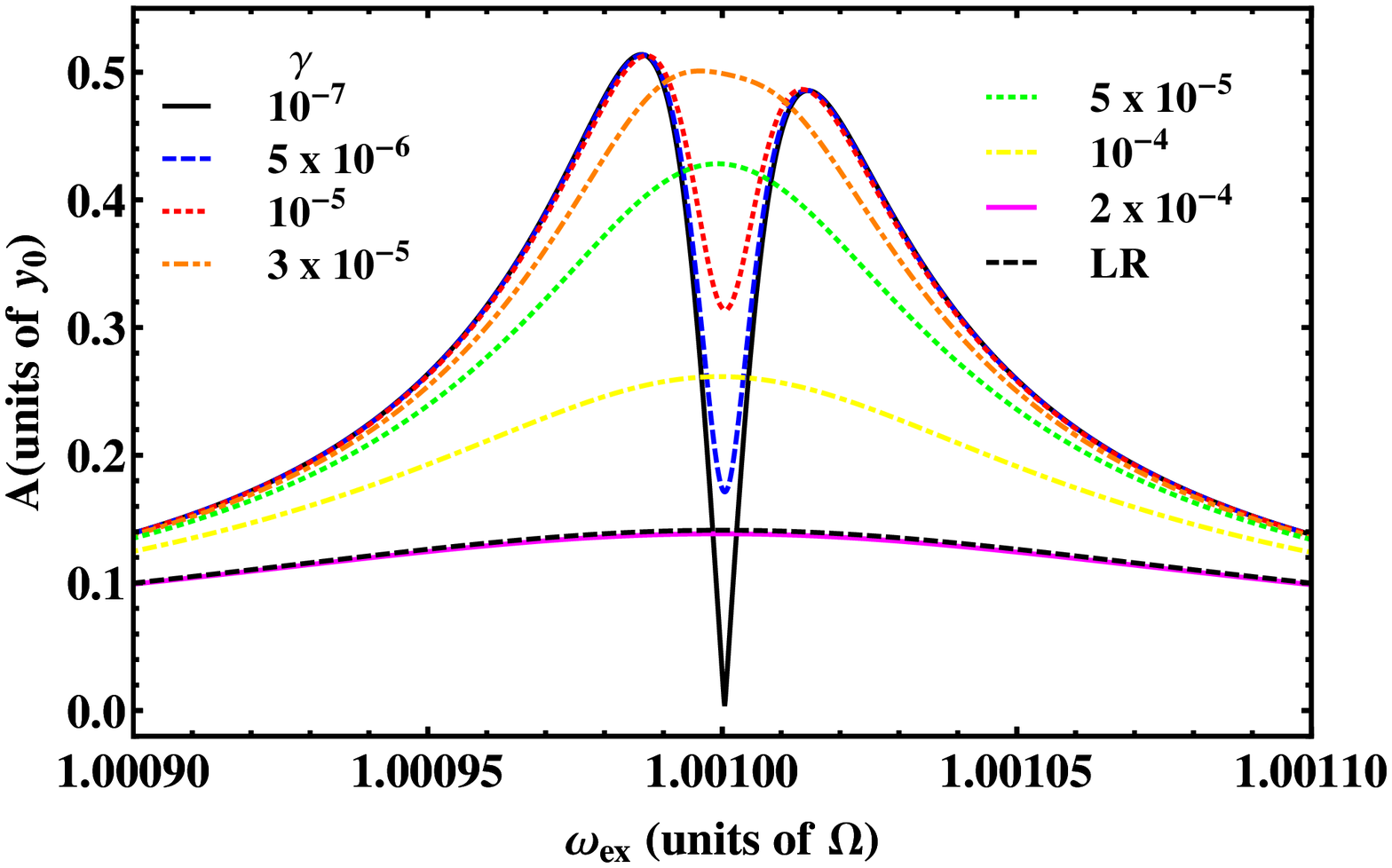}
\caption{Amplitude $A$ versus driving frequency at low temperature $k_B T/(\hbar\Omega)=10^{-3}$ for different damping strengths $\gamma$. For the rest of the parameters we take $y_0^4\alpha/(\hbar\Omega)=4/3\cdot 10^{-3}$, $y_0F/(\hbar\Omega)=2^{3/2}\cdot10^{-5}$, $\Omega=1$ and varying damping: $\gamma/\Omega=10^{-7},5\cdot10^{-6},1\cdot10^{-5},3\cdot10^{-5},5\cdot10^{-5},1\cdot10^{-4},2\cdot10^{-4}$.\label{A}}
\end{figure}

We compute the amplitude in the low temperature regime, for fixed driving
amplitude and varying damping strengths 
by solving Eq. (\ref{eqsys}) for the one-photon resonance $E_{0,0}=E_{1,1}$ . Depending on the actual value of the damping, introduced by the bath, either an antiresonant behaviour, at small damping, or a resonant one, at large damping, is observed. The resonant/ antiresonant lineshape depends on the ratio of damping and minimal splitting: $\hbar\sqrt{2}\gamma/\left[1-\frac{3}{8\hbar\Omega}\alpha\right] F y_0$.  In case of high damping, the amplitude is very small and broad showing resonant behaviour, where the asymmetry is smeared out completly. If we decrease the damping, the amplitude increases and the width shrinks until we reach a critical value for the damping. This critical value occurs when damping and minimal splitting are almost equal. Lowering the damping below the critical value, a cusp-like profile arises: the antiresonance. The treatment of higher orders in the driving and the interplay of driving and nonlinearity introduce additionally an asymmetry in the response with respect to $\omega_{ex}=\Omega+\frac{3}{4\hbar}\alpha y_0^4$. For very large damping we observe the corresponding linear response (LR) of a linear oscillator with eigenfrequency $\Omega+\frac{3}{4\hbar}\alpha y_0^4$, indicated by the black dashed line in Figure \ref{A}. 
So we can make the connection to the linear response of a driven damped harmonic oscillator, which is resonant at the frequency $\omega_{ex}=\Omega+\frac{3}{4\hbar}\alpha y_0^4$, if considering first order perturbation theory in the nonlinearity.
\subsection{Phase}
The phase for the one photon resonance is given in Figure \ref{phase}:
\begin{figure}[h!]
\includegraphics[width=12.0cm]{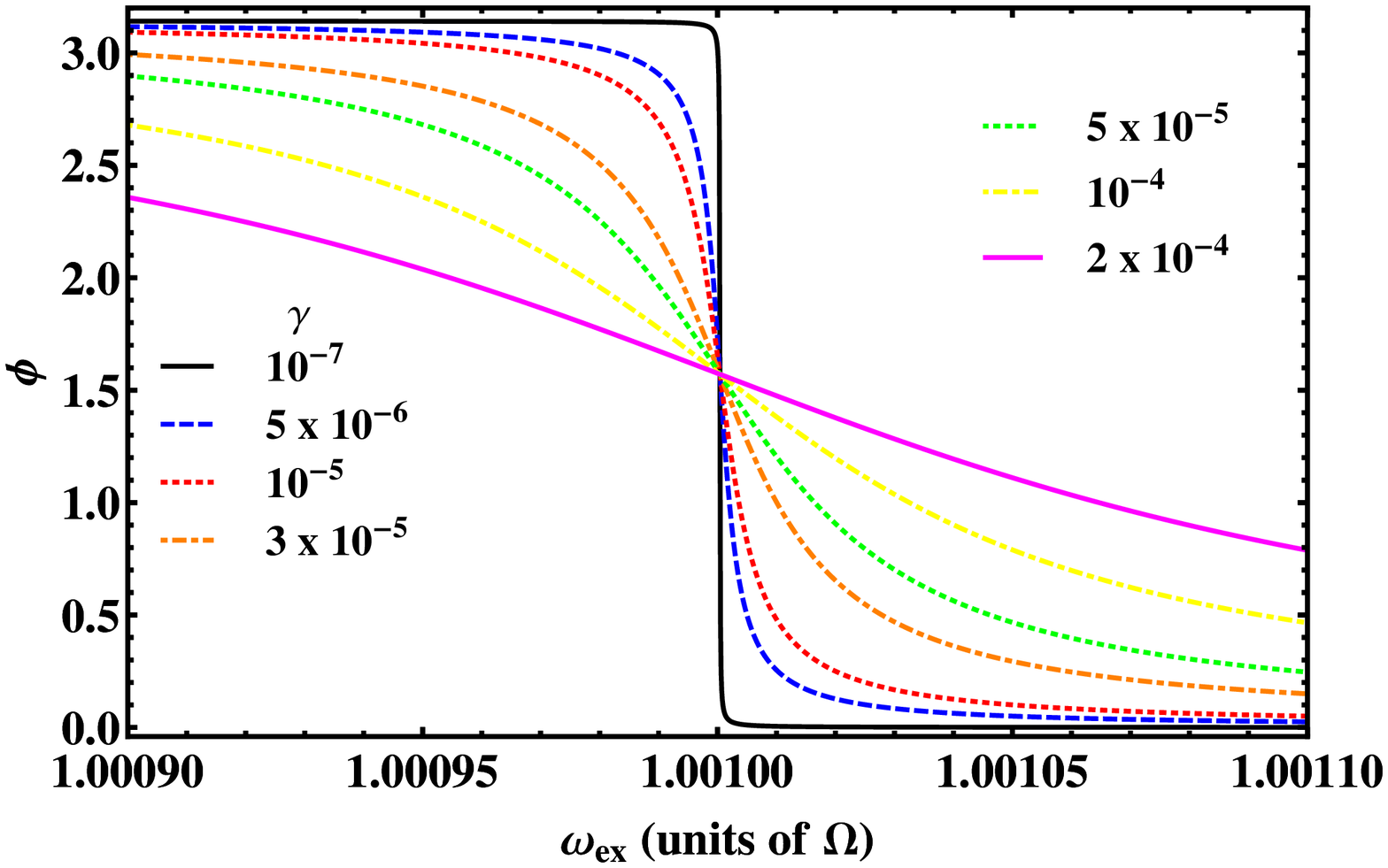}
\caption{Phase $\phi$ versus driving fequency at low temperature $k_B T/(\hbar\Omega)=10^{-3}$ for different damping strengths $\gamma$. For the rest of the parameters we take $\alpha=4/3\cdot 10^{-3}$, $F=2^{3/2}\cdot10^{-5}$, $\Omega=1$ and varying damping: $\gamma=10^{-7},5\cdot10^{-6},1\cdot10^{-5},3\cdot10^{-5},5\cdot10^{-5},1\cdot10^{-4},2\cdot10^{-4}$.\label{phase}}
\end{figure}
It jumps by $\pi$ at the one-photon resonance and the step becomes smoother the larger the damping is.
\section{Conclusions}\label{Conclusions}
In this work we have discussed two perturbative approaches, App I and App II, to calculate the Floquet quasienergies and states of the quantum Duffing oscillator beyond a RWA approach. Additionally, the stationary dynamics of the expectation value of the oscillator position was obtained. Specifically, analytical results where derived \textit{off} and near resonance with App I and \textit{at} the one-photon resonance within App II. For App II, based on Van Vleck perturbation theory, we also assumed that the driving is much weaker than the nonlinearity. We showed that in the parameter regime near resonance the Van Vleck approach recovers the results of App I based on the exact Floquet states of the driven linear oscillator, for both the eigenfrequencies in second order in the driving strength $F$ and the matrix elements of the position operator in the Floquet basis to first order in $F$. The comparison allows to treat the quantum Duffing oscillator for arbitrary frequencies, as well as an estimation of how good the Van Vleck approach is as App I is exact to all orders in the driving. As an application of our formalism we considered the dynamics of a quantum Duffing oscillator coupled to an Ohmic bath and calculated its response near an one-photon resonance. Dissipative effects strongly affect the behaviour of the Duffing oscillator in the resonance region. We observe upon variation of the damping strength a transition from antiresonant to resonant behaviour and that the shape of the antiresonance with respect to the one-photon resonance condition is asymmetric.

\section{Acknowledgements}
We thank P. H\"anggi for numerous interesting discussions on the physics of driven quantum systems along the years and his continuous support. We acknowledge the financial support under the DFG programs SFB 631 and GRK 638. Further we would like to thank Johannes Hausinger for many helpful discussions.

 \appendix
\section{Floquet theory}\label{FloqTh}
The Floquet theorem states that a Schr\"odinger equation involving a time-periodic Hamiltonian ($\hat{H}(t)=\hat{H}(t+T_{\omega_{ex}})$ with period $T_{\omega_{ex}}=2\pi/\omega_{ex}$) \cite{Floquet,Shirley,Sambe,Grifoni304,Kohler2}:
\begin{eqnarray}\label{timeSGL}
i\hbar\partial_t\ket{\psi(t)}=\hat{H}(t)\ket{\psi(t)}
\end{eqnarray}
has a complete set of solutions:
\begin{eqnarray}
|\psi_j(t)\rangle=\exp(-i\epsilon_j t/\hbar)|\phi_j (t)\rangle,
\end{eqnarray}
where $|\phi_j (t)\rangle=|\phi_j(t+T_{\omega_{ex}})\rangle$. The quasienergies $\epsilon_j$ and Floquet states $|\phi_j (t)\rangle$ solve the eigenvalue equation of the Floquet Hamiltonian:
\begin{eqnarray}
\label{q11}
\hat{\mathcal{H}}(t) |\phi_j (t)\rangle=\left[\hat{H}(t)-i\hbar\frac{\partial}{\partial t}\right]|\phi_j (t)\rangle=\epsilon_j|\phi_j (t)\rangle.
\end{eqnarray}
Defining $\ket{\phi_{j,n}(t)}=\exp(-in\omega_{ex}t)\ket{\phi_j(t)}$ and inserting into Eq. (\ref{q11}) we find that 
$\ket{\phi_{j,n}(t)}$ is also an eigenstate of the Floquet Hamiltonian, but with the eigenvalue $\epsilon_{j,n}=\epsilon_j-n\hbar\omega_{ex}$ differing by a multiple integer of $\hbar\omega_{ex}$. These Floquet states are therefore physically identical. In other words the spectrum of the Floquet Hamiltonian has a Brillouin zone structure, each Brillouin zone being of size $\omega_{ex}$. To find a complete set of solutions $\{|\psi_j(t)\rangle\}$, it is sufficient to consider only those Floquet states and quasienergies which lie within a single Brillouin zone, i.e. $-\hbar\omega_{ex}/2\leq\epsilon_j<\hbar\omega_{ex}/2$. Moreover $\epsilon_j\equiv\epsilon_{j,0}$ and $\ket{\phi_j(t)}:=\ket{\phi_{j,0}(t)}$. 
The eigenstates $|\psi_j(t)\rangle$ as well as the Floquet states $|\phi_j (t)\rangle$ are elements of the Hilbert space $\mathcal{R}$. For convenience we also define $\mathcal{T}$ as the space of $T_{\omega_{ex}}$-periodic functions with the inner product: 
\begin{eqnarray}\label{innprod}
(f,g)&=&\frac{1}{T_{\omega_{ex}}}\int_0^{T_{\omega_{ex}}}dt f^{*}(t)g(t).
\end{eqnarray}
An orthonormalized basis of $\mathcal{T}$ is given by the functions: $\varphi_n(t)=\exp(-in\omega_{ex}t)$, $n$ integer. The basis set $\{\varphi_n\}$ is orthonormalized and complete:
\begin{eqnarray}
(\varphi_n,\varphi_n')&=&\delta_{n,n'},\\
\frac{1}{T_{\omega_{ex}}}\sum_n\varphi_n^*(t)\varphi_n(t')&=&\delta_{T_{\omega_{ex}}}(t-t'),
\end{eqnarray}
where $\delta_{T_{\omega_{ex}}}$ is the $T_{\omega_{ex}}$-periodic delta function. The scalar product in the composite Hilbert space $\mathcal{R}\otimes\mathcal{T}$ is then given by:
 \begin{eqnarray}\label{Fl2}
 \langle\langle \phi_j|\phi_k\rangle\rangle&:=&\frac{1}{T_{\omega_{ex}}}\int_0^{T_{\omega_{ex}}}dt \langle \phi_j(t)|\phi_k(t)\rangle.
 \end{eqnarray}
The decomposition of $\ket{\phi_j(t)}$ into basis functions $\varphi_n(t)$ is equivalent to an expansion in Fourier series:
  \begin{eqnarray}\label{Fl1}
 \ket{\phi_{j,n}(t)}&=&\sum_l\exp(-il\omega_{ex}t)\ket{\phi_j^{(l-n)}},\\
 \ket{\phi_j^{(n)}}&=&\frac{1}{T_{\omega_{ex}}}\int_0^{T_{\omega_{ex}}}dt\exp(in\omega_{ex}t)\ket{\phi_j(t)}\\
&=&\frac{1}{T_{\omega_{ex}}}\int_0^{T_{\omega_{ex}}}dt\ket{\phi_{j,-n}(t)}.\nonumber
 \end{eqnarray}
For a basis independent notation we introduce the state vectors $|n)$ with $(t|n):=\varphi_n(t)$.
Then, in the composite Hilbert space $\mathcal{R}\otimes\mathcal{T}$ we define the state:
\begin{equation}\label{double}
\ket{\phi_{j,n}}\rangle=\sum_l\ket{\phi_j^{(l-n)}}\otimes |l),
\end{equation}
and $\ket{\phi_{j,n}(t)}=(t\ket{\phi_{j,n}}\rangle.$
In particular, due to the orthogonality of the Floquet states $\ket{\phi_{j,m}(t)}$, it also holds the orthonormality relation:
\begin{eqnarray}
\langle\braket{\phi_{i,n}}{\phi_{j,m}}\rangle=\frac{1}{T_{\omega_{ex}}}\int_0^{T_{\omega_{ex}}}dt\braket{\phi_{i,n}(t)}{\phi_{j,m}(t)}=\delta_{ij}\delta_{nm}.
\end{eqnarray}
Additionally we define:
\begin{eqnarray}\label{hmatrix}
\mathcal{H}_{ij}^{nm}&\equiv&\langle\bra{\phi_{i,n}}\hat{\mathcal{H}}\ket{\phi_{j,m}}\rangle=\frac{1}{T_{\omega_{ex}}}\int_0^{T_{\omega_{ex}}}dt\bra{\phi_{i,n}(t)}\hat{\mathcal{H}}(t)\ket{\phi_{j,m}(t)}.
\end{eqnarray}
Hence through the expansion of the Hilbert space it is possible to treat the time-dependent problem (\ref{timeSGL}) as a time-independent one by expressing the Floquet Hamiltonian in a basis of $\mathcal{R}\otimes\mathcal{T}$. Which basis of $\mathcal{R}\otimes\mathcal{T}$ is the most convenient to express $\mathcal{H}$ depends on the specific problem. For example in the basis $\{\ket{\phi_{j,n}}\rangle\}$ of the Floquet states the eigenvalue equation reads:
\begin{eqnarray}\label{heigenvalue}
\hat{\mathcal{H}}\ket{\phi_{j,m}}\rangle=\epsilon_{j,m}\ket{\phi_{j,m}}\rangle.
\end{eqnarray}
Equivalently, using the expansion (\ref{double}) it also follows from (\ref{hmatrix}) and (\ref{heigenvalue})
\begin{eqnarray}\label{fouriereigenvalue}
\sum_{l'}\left(\hat{H}^{(l-l')}-l'\hbar\omega_{ex}\delta_{ll'}\right)\ket{\phi_j^{(l'-m)}}=\epsilon_{j,m}\ket{\phi_j^{(l-m)}},
\end{eqnarray}
where $\hat{H}^{(l)}$ are the Fourier components in the Fourier expansion of $\hat{H}(t)$.\\
The eigenenergies and eigenstates of $\hat{\mathcal{H}}$ are known only in very few cases, among which the case of the driven linear oscillator, see \ref{FloqLO}. For a generic time-periodic Hamiltonian only approximated solutions of Eqs. (\ref{hmatrix}) or (\ref{heigenvalue}) can be found. Two complementary approximation schemes are discussed for the Duffing oscillator case in section \ref{QDOmodel}.
\section{Floquet states of the driven harmonic oscillator}
\label{FloqLO}
Following Husimi, Perelomov and Breuer \cite{Husimi,Popov,Breuer} the quasienergy spectrum can be determined exactly for a periodically driven harmonic oscillator. The solution $\braket{y}{\psi_j(t)}=\psi_j(y,t)$ of (\ref{q11}) for the time-dependent Hamiltonian $\hat{H}_{\rm LO}(t)$, see Eq. (\ref{App1}), is:
\begin{subequations}
\begin{eqnarray}
\psi_j(y,t)&=&\overline{\phi_j}(y-\xi(t))\exp\left\{-i\left[\left(j+\frac{1}{2}\right)\Omega t-\frac{1}{2\hbar}\int_0^t f(t')\xi(t')dt'\right]\right\}\nonumber\\
&\equiv&\phi_j^{(0)}(y,t)\exp(-i\epsilon_j^{(0)}t/\hbar)\label{flo1a},\\
\overline{\phi}_j(y)&=&\braket{y}{j}_0=\frac{1}{\sqrt{y_0}}\frac{1}{\sqrt{2^jj!\sqrt{\pi}}}\exp\left(-\frac{y^2}{2y_0^2}\right)H_j\left(\frac{y}{y_0}\right),\label{flo1b}
\end{eqnarray}
\end{subequations}
with $\ket{j}_0$ the eigenstates of the linear, undriven oscillator and $H_j\left(\frac{y}{y_0}\right)$ the Hermite polynomials.
Finally $\xi(t)$ is the steady state solution of the corresponding classical equation,
\begin{eqnarray}
M\ddot{\xi}(t)+M\Omega^2\xi(t)&=&f(t),
\end{eqnarray}
which is for a driving of the form $f(t)=-F\cos(\omega_{ex} t)$:
\begin{eqnarray}
\xi(t)&=&\frac{F}{M(\omega_{ex}^2-\Omega^2)}\cos(\omega_{ex}t).
\end{eqnarray}
The quasienergy spectrum for the harmonic oscillator is for a cosine-like driving term:
\begin{eqnarray}\label{flo5}
\epsilon_j^{(0)}=\epsilon_{j,0}^{(0)}&=&\hbar\Omega\left(j+\frac{1}{2}\right)+\frac{F^2}{4M(\omega_{ex}^2-\Omega^2)}.
\end{eqnarray}
Finally,
\begin{eqnarray}
\phi_j^{(0)}(y,t)&=&\overline{\phi}_j(y-\xi(t))\exp(-i\theta(F^2,t))
\end{eqnarray}
with
\begin{eqnarray}\label{drivingphase}
\theta(F^2,t)&=&-\frac{1}{2\hbar}\left[\int_0^tdt' f(t')\xi(t')-\frac{t}{T_{\omega_{ex}}}\int_0^{T_{\omega_{ex}}}dt' f(t')\xi(t')\right].
\end{eqnarray}
\section{Fourier components}\label{Fouriercomponents}
In this appendix we evaluate the matrix elements of the perturbation $\hat{V}_\alpha$ on the Floquet basis of the driven linear oscillator as well as the Fourier coefficients $v_{kj}^{(n)}$.
\begin{eqnarray}
_0\langle\phi_k(t)|\hat{V}_\alpha|\phi_j(t)\rangle_0&=&\frac{\alpha}{4}\int_{-\infty}^\infty dy'dy''\ _0\langle\phi_k(t)|y'\rangle\langle y'|\hat{y}^4|y''\rangle\langle y''|\phi_j(t)\rangle_0\\
&=&\frac{\alpha}{4}\int_{-\infty}^\infty dy\overline{\phi}_k(y)\left[y+\xi(t)\right]^4\overline{\phi}_j(y)\nonumber\\
&=&\frac{\alpha}{4}\int_{-\infty}^\infty dy\overline{\phi}_k(y)\left[y^4+4\xi(t)y^3+6\xi(t)^2y^2\right.\nonumber\\
&&\left.+4\xi(t)^3y+\xi(t)^4\right]\overline{\phi}_j(y)\nonumber\\
&=&\sum_n\exp(-in\omega_{ex}t)\underbrace{_0\langle\phi_k(t)|\hat{V}_\alpha|\phi_j(t)\rangle_0^{(n)}}_{\equiv v_{kj}^{(n)}}.
\end{eqnarray}

The Fourier coefficients $v_{kj}^{(n)}$ are given, using the notation $A_\xi\equiv\frac{F}{M(\omega_{ex}^2-\Omega^2)}$, by:
\begin{eqnarray}
v_{kj}^{(0)}&=&\frac{\alpha}{4}\left[\delta_{kj}\left(\frac{3}{2}(2j+1)y_0^2A_\xi^2+\frac{3}{2}(j(j+1)+\frac{1}{2})y_0^4+\frac{3}{8}A_\xi^4\right)\right.\\
&&+\delta_{k,j+2}\left(\frac{3}{2}y_0^2A_\xi^2+y_0^4(j+\frac{3}{2})\right)\sqrt{(j+1)(j+2)}\nonumber\\
&&+\delta_{k,j-2}\left(\frac{3}{2}y_0^2A_\xi^2+y_0^4(j-\frac{1}{2})\right)\sqrt{j(j-1)}\nonumber\\
&&+\delta_{k,j+4}\frac{y_0^4}{4}\sqrt{(j+1)(j+2)(j+3)(j+4)}\nonumber\\
&&+\delta_{k,j-4}\frac{y_0^4}{4}\sqrt{j(j-1)(j-2)(j-3)}\left.
\right]\nonumber,
\end{eqnarray}
\begin{eqnarray}
v_{kj}^{(\pm 1)}&=&\frac{\alpha}{4}\left[\delta_{k,j+1}\left(\frac{3}{2\sqrt{2}}\sqrt{j+1}y_0A_\xi^3+\frac{3!\sqrt{2}}{4}(j+1)\sqrt{j+1}A_\xi y_0^3\right)\right.\\
&&+\delta_{k,j-1}\left(\frac{3}{2\sqrt{2}}\sqrt{j}y_0A_\xi^3+\frac{3!\sqrt{2}}{4}j\sqrt{j}A_\xi y_0^3\right)\nonumber\\
&&+\delta_{k,j+3}\sqrt{(j+3)(j+2)(j+1)}\frac{2^{3/2}}{4}y_0^3A_\xi\nonumber\\
&&+\delta_{k,j-3}\sqrt{j(j-1)(j-2)}\frac{2^{3/2}}{4}y_0^3A_\xi\left.
\right]\nonumber,
\end{eqnarray}
\begin{eqnarray}
v_{kj}^{(\pm 2)}&=&\frac{\alpha}{4}\left[\delta_{kj}\left(\frac{3}{4}(2j+1)y_0^2A_\xi^2+\frac{1}{4}A_\xi^4\right)\right.\\
&&+\delta_{k,j+2}\frac{3}{4}y_0^2A_\xi^2\sqrt{(j+1)(j+2)}\nonumber\\
&&+\delta_{k,j-2}\frac{3}{4}y_0^2A_\xi^2\sqrt{j(j-1)}\left.
\right]\nonumber,
\end{eqnarray}
\begin{eqnarray}
v_{kj}^{(\pm 3)}&=&\frac{\alpha}{4}\left[\delta_{k,j+1}\frac{1}{2\sqrt{2}}\sqrt{j+1}y_0A_\xi^3\right.\\
&&\left.+\delta_{k,j-1}\frac{1}{2\sqrt{2}}\sqrt{j}y_0A_\xi^3\right]\nonumber,
\end{eqnarray}
\begin{eqnarray}
v_{kj}^{(\pm 4)}&=&\frac{\alpha}{4}\left[\frac{1}{16}A_\xi^4\right]\delta_{kj}.
\end{eqnarray}
\section{Van Vleck perturbation theory}\label{AppVVK}
In the following we give a basic introduction of the Van Vleck perturbation theory \cite{Shavitt1980,Cohen1992,Certain,Kirtman}.
It allows to calculate eigenenergies and eigenfunctions of Hamiltonians $\hat{H}$ whose spectrum is splitted into well-defined manifolds (denoted by Greek indices) \cite{Cohen1992,Certain,Kirtman,Certain2,Aravind}. Within Van Vleck perturbation theory an effective Hamiltonian $\hat{H}_{\rm eff}=\exp(i\hat{S})\hat{H}\exp(-i\hat{S})$ is constructed whose spectrum is the same as that of the original Hamiltonian
\begin{eqnarray}
\hat{H}&=&\hat{H}^{(0)}+\hat{V},
\end{eqnarray}
but only connects almost degenerate levels within a given manifold. The eigenstates are then calculated from the eigenstates of the effective Hamiltonian via a back transformation. For time-dependent Hamiltonians one introduces an effective Floquet Hamiltonian $\hat{\mathcal{H}}_{\rm eff}$ in the composite Hilbert space $\mathcal{R}\otimes\mathcal{T}$. The effective quasienergies up to second order in the perturbation are determined by:
\begin{eqnarray}
&&\langle\bra{i,\alpha}\hat{\mathcal{H}}_{\rm eff}\ket{j,\alpha}\rangle=E_{i,\alpha}\delta_{ij}+\langle\bra{i,\alpha}\hat{V}\ket{j,\alpha}\rangle\\
&&+\frac{1}{2}\sum_{k,\gamma\neq\alpha}\langle\bra{i,\alpha}\hat{V}\ket{k,\gamma}\rangle\langle\bra{k,\gamma}\hat{V}\ket{j,\alpha}\rangle\left[\frac{1}{E_{i,\alpha}-E_{k,\gamma}}+\frac{1}{E_{j,\alpha}-E_{k,\gamma}}\right]\nonumber\\
&&=E_{i,\alpha}\delta_{ij}+\langle\bra{i,\alpha}\hat{V}\ket{j,\alpha}\rangle+\langle\langle i,\alpha|(\hat{R}\hat{V})^\dagger\hat{V}|j,\alpha\rangle\rangle\nonumber\\
&&+\langle\langle i,\alpha|\hat{V}(\hat{R}\hat{V})|j,\alpha\rangle\rangle,\nonumber
\end{eqnarray}
where we introduced the reduced resolvent:
\begin{eqnarray}
\hat{R}&=&\sum_{k,\gamma}\ ' |k,\gamma\rangle\rangle\langle\langle k,\gamma|/(E-E_{k,\gamma}).
\end{eqnarray}
The prime over the sum denotes that all the states belonging to the manifold $\alpha$ under consideration are excluded from the sum.
The states of the original Floquet Hamiltonian $\hat{\mathcal{H}}$ are given by applying a back transformation connecting different manifolds: $\exp(-i\hat{S})\ket{j,\alpha}\rangle=(1-i\hat{S}^{(1)}-i\hat{S}^{(2)}+i\hat{S}^{(1)}i\hat{S}^{(1)}/2+\dots)\ket{j,\alpha}\rangle$, where:
\begin{eqnarray}
\langle\bra{i,\alpha}i\hat{S}^{(1)}\ket{j,\beta}\rangle&=&\frac{\langle\bra{i,\alpha}\hat{V}\ket{j,\beta}\rangle}{E_{i,\alpha}-E_{j,\beta}}\\
\langle\bra{i,\alpha}i\hat{S}^{(2)}\ket{j,\beta}\rangle&=&\frac{\langle\bra{i,\alpha}\hat{V}\ket{k,\gamma}\rangle\langle\bra{k,\gamma}\hat{V}\ket{j,\beta}\rangle}{2(E_{j,\beta}-E_{i,\alpha})}\left[\frac{1}{E_{k,\gamma}-E_{i,\alpha}}+\frac{1}{E_{k,\gamma}-E_{j,\beta}}\right]\nonumber\\&&
+\sum_k\frac{1}{E_{j,\beta}-E_{i,\alpha}}\frac{\langle\bra{i,\alpha}\hat{V}\ket{k,\beta}\rangle\langle\bra{k,\beta}\hat{V}\ket{j,\beta}\rangle}{E_{k,\beta}-E_{i,\alpha}\nonumber}\\&&
+\sum_k\frac{1}{E_{j,\beta}-E_{i,\alpha}}\frac{\langle\bra{i,\alpha}\hat{V}\ket{k,\alpha}\rangle\langle\bra{k,\alpha}\hat{V}\ket{j,\beta}\rangle}{E_{k,\alpha}-E_{j,\beta}}\nonumber.
\end{eqnarray}
The construction using the reduced resolvent is more easily comparable to conventional degenerate perturbation theory \cite{Nolting,Stone}. After identification of the degenerate levels the modifications to the eigenvectors are given by calculating all possible matrix elements from the degenerate levels out of the manifold. 
\section{Comparison for the states}\label{Compstates}
As seen in section \ref{Compquasien}, App II yields away from resonance the same quasienergies as App I expanded up to second order in the driving amplitude $F$. In the following we determine whether the states behave in the same way. For simplicity we compare the two approaches for the case of the driven linear oscillator. This corresponds in the perturbative approach, App II, to expand to zeroth order in the nonlinearity and first order in the driving.
\subsection{Floquet states in App I}
According to equation (\ref{flo1a}) the Floquet states of the driven linear oscillator $\{\ket{\phi_j(t)}_0\}$ can be obtained from those of the undriven linear oscillator $\{\ket{j}_0\}$ by applying a time-dependent translation:
\begin{eqnarray}
\hat{U}(\xi(t))&=&\exp\left(-\frac{\xi(t)}{y_0\sqrt{2}}(\hat{a}-\hat{a}^\dagger)\right),
\end{eqnarray}
and accounting for a driving dependent phase $\exp(-i\theta(F^2,t))$, see Eq. (\ref{drivingphase}). Here $\hat{a}^\dagger$ and $\hat{a}$ are the usual creation/annihilation operator related to the linear oscillator.
To compare to the results of App II we give the Floquet states from App I up to first order in the driving amplitude and zeroth in the nonlinearity
\begin{eqnarray}
\hat{U}(\xi(t))\exp(-i\theta(F^2,t))&=&1+\frac{\xi(t)}{y_0\sqrt{2}}(\hat{a}^\dagger-\hat{a})+\mathcal{O}(F^2),
\end{eqnarray}
\begin{eqnarray}
\ket{\phi_j(t)}_0&=&\hat{U}(\xi(t))\exp(-i\theta(F^2,t))\ket{j}_0\\
&=&\ket{j}_0+\frac{\xi(t)}{y_0\sqrt{2}}\left(\sqrt{j+1}\ket{j+1}_0-\sqrt{j}\ket{j-1}_0\right)+\mathcal{O}(F^2)\nonumber.
\end{eqnarray}
In the composite Hilbert space $\mathcal{R}\otimes \mathcal{T}$ the Floquet state corresponding to $\epsilon_{j,0}^{(0)}$ is:
\begin{eqnarray}\label{statelin}
\ket{\phi_{j,0}}\rangle_0&=&\ket{j,0}\rangle_0+\frac{F}{2\sqrt{2}My_0(\omega_{ex}^2-\Omega^2)}
\left(\sqrt{j+1}\ket{j+1,-1}\rangle_0\right.\\
&&\left.+\sqrt{j+1}\ket{j+1,+1}\rangle_0-\sqrt{j}\ket{j-1,-1}\rangle_0-\sqrt{j}\ket{j-1,+1}\rangle_0\right).\nonumber
\end{eqnarray}

\subsection{Floquet states in App II}
In App II the counterpart of Eq. (\ref{statelin}) is the state:
\begin{eqnarray}
\ket{-_{j,0}}\rangle_0
&=&\ket{j,0}\rangle_0+\frac{F}{2\sqrt{2}My_0\Omega }\left(\frac{\sqrt{j+1}\ket{j+1,+1}\rangle_0}{\omega_{ex}-\Omega}-
\frac{\sqrt{j+1}\ket{j+1,-1}\rangle_0}{\omega_{ex}+\Omega}\right.\nonumber\\&&\left.
-\frac{\sqrt{j}\ket{j-1,-1}\rangle_0}{\omega_{ex}-\Omega}
+\frac{\sqrt{j}\ket{j-1,+1}\rangle_0}{\omega_{ex}+\Omega}\nonumber
\right).
\end{eqnarray}
Note that by reducing to the linear case not only the energies $E_{j,0}$ and $E_{j+1,+1}$ but also $E_{j-1,-1}$ are equal if $\Omega\approx\omega_{ex}$. These are the dominant contributions, as $1/(\Omega-\omega_{ex})>>1$ if $\Omega\approx\omega_{ex}$:
\begin{equation}\label{dom}
\ket{-_{j,0}}\rangle_0
\cong\ket{j,0}\rangle_0+\frac{F}{2\sqrt{2}My_0\Omega }\left(\frac{\sqrt{j+1}\ket{j+1,+1}\rangle_0}{\omega_{ex}-\Omega}
-\frac{\sqrt{j}\ket{j-1,-1}\rangle_0}{\omega_{ex}-\Omega}
\right).
\end{equation}
\subsubsection{Comparison}
By comparing the states we see that the exact linear oscillator Floquet states from App I 
are proportional to $\xi(t)$ and hence are obtained as a linear combination of rotating and anti-rotating contributions. This accounts for the fact that the spectrum of the underlying oscillator is equidistant. Consequently the driving can excite both upwards and downwards transitions $\ket{j}_0\rightarrow\ket{j+1}_0$, $\ket{j}_0\rightarrow\ket{j-1}_0$ with the same weight. In contrast, App II in a vicinity of a $N$-photon resonance 
includes rotating and antirotating terms but with different weights. In fact, by choosing a resonance condition certain transitions are preferred, namely those within the resonant levels and those lying closest to these, giving the dominant contributions (see Eq. \ref{dom}). The actual form for the states in App II is determined by the structure of the quasienergy-spectrum, i.e., by the resonance condition determining the structure of the manifold. As the manifold structure of App II is crucially depending on a finite nonlinearity to obtain a doublet, a reduction to the linear case is not possible. It would destroy the ordering of the manifolds, as not only $E_{j,0}$ and $E_{j+1,1}$ but also $E_{j+n,n}$, $n>1$ and $E_{j-n,-n}$, $n\geq1$, become degenerate. The way of treating the Floquet Hamiltonian thus results in different symmetry properties and weights for the corresponding states. 







\end{document}